\RequirePackage{lineno}
\documentclass[twocolumn,showpacs,superscriptaddress,amsmath,amssymb,nofootinbib]{revtex4-2}
\usepackage{color}
\usepackage{float}
\usepackage{graphicx}
\usepackage{longtable}
\usepackage{subfigure}
\usepackage{epsfig}
\usepackage{overpic}
\usepackage{dcolumn}
\usepackage{ulem}
\usepackage{bm}
\usepackage{lineno}
\usepackage{xspace}
\usepackage{multirow}
\usepackage{epstopdf}
\usepackage{xcolor}
\usepackage{soul}
\usepackage{verbatim}
\usepackage{enumitem}
\usepackage{todonotes}
\usepackage[toc,page]{appendix}
\usepackage[colorlinks,allcolors=black]{hyperref}
\usepackage{subfigure}
\usepackage{diagbox}

\newcommand{\moe}{\affiliation{Key Laboratory of Atomic and Subatomic Structure and Quantum Control (MOE), Guangdong Basic Research Center of Excellence for Structure and Fundamental Interactions of Matter, Institute of Quantum Matter, South China Normal University, Guangzhou 510006, China
}}

\newcommand{\iqm}{\affiliation{Guangdong-Hong Kong Joint Laboratory of Quantum Matter, Guangdong Provincial Key Laboratory of Nuclear Science, Southern Nuclear Science Computing Center, South China Normal University, Guangzhou 510006, China}}

\newcommand{\scnt}{\affiliation{Southern Center for Nuclear-Science Theory (SCNT), Institute of Modern Physics, Chinese Academy of Sciences, Huizhou 516000, Guangdong Province, China}}

\begin{document}
\include{def-com}
\title{\boldmath Two-body hadronic weak decays of bottomed hadrons}

\author {Ying Zhang}\email{yingzhang@m.scnu.edu.cn}
\moe\iqm

\author{Guangzhao He}
\moe\iqm

\author{Quanxing Ye}
\moe\iqm

\author{Da-Cheng Yan}\email{yandac@126.com}
\affiliation{School of Mathematics and Physics, Changzhou University, Changzhou, Jiangsu 213164, China}

\author{Jun Hua}\email{junhua@m.scnu.edu.cn}
\moe\iqm

\author{Qian Wang}\email{qianwang@m.scnu.edu.cn}
\moe\iqm\scnt 

\date{\today}

\begin{abstract}
The structure of light diquarks plays a crucial role in the formation of exotic hadrons beyond the conventional quark model, especially in their line shapes of bottomed hadron decays.
We study the two-body hadronic weak decays of bottomed baryons 
and bottomed mesons to probe the light diquark structure and pin down the quark-quark correlations in the diquark picture.  
We find that the light diquark does not favor a compact structure. For instance, the isoscalar diquark $[ud]$ in $\Lambda_{b}^{0}$  can be easily split and rearranged to form $\Sigma_{c}^{(*)}\bar{D}^{(*)}$ via the color-suppressed transition. This provides a hint that the hidden charm pentaquark states produced in $\Lambda^0_b$ decays could be the $\Sigma_{c}^{(*)}\bar{D}^{(*)}$ hadronic molecular candidates.
 This quantitative study resolves the apparent conflicts between the production mechanism and molecular nature of these $P_c$ states observed in experiment.
\end{abstract}

\maketitle

\section{Introduction}
 Although the success of the conventional quark model indicates that meson is made of a pair of quark and anti-quark and baryon is made of three quarks~\cite{GellMann:2019ASM,Zweig:1964ruk}, the strong interaction dynamics, namely Quantum Chromo-dynamics (QCD) does not eliminate the existence of more profound objects beyond the hadrons in the conventional quark model picture. Since the observation of $X(3872)$ in 2003~\cite{Belle:2003nnu}, numerous exotic candidates which do not fit into the conventional quark model spectrum have been observed in experiment, 
such as the $Z(3930)$~\cite{Belle:2005rte}, the $Y(4140)$~\cite{CDF:2009jgo}, the $X(4500)$~\cite{LHCb:2016axx}, the $P_c(4312)$, $P_c(4440)$, $P_c(4457)$~\cite{LHCb:2019kea} and so on. Among these observations the hidden charm pentaquarks, $P_c(4312)$, $P_c(4440)$, and $P_c(4457)$ observed by the LHCb collaboration~\cite{LHCb:2019kea}, are ideal hadronic molecular candidates for $\Sigma_{c}^{(*)}\bar{D}^{(*)}$ since their masses are close to the $\Sigma_{c}^{(*)}\bar{D}^{(*)}$ threshold and the $\Sigma_{c}^{(*)}\bar{D}^{(*)}$ interactions turn out to be attractive near threshold~\cite{Wu:2010jy,Wang:2011rga,Wu:2012md,Xiao:2013yca,Karliner:2015ina,Liu:2019tjn,Du:2019pij,Peng:2022iez,Yang:2022ezl,Xing:2022ijm,Yalikun:2021bfm,Yan:2021nio,Du:2021fmf,Kuang:2020bnk,Pan:2019skd,Liu:2019zvb,Wang:2019nwt,Chen:2019asm,Wang:2019ato,Chen:2019bip,Guo:2019fdo, He:2019ify, Guo:2019kdc, Shimizu:2019ptd, Xiao:2019mvs, Xiao:2019aya,   Meng:2019ilv, Wu:2019adv, Xiao:2019gjd, Voloshin:2019aut, Sakai:2019qph, Yamaguchi:2019seo, Lin:2019qiv, Gutsche:2019mkg, Burns:2019iih}. 
However, whether these hidden charm pentaquarks behave as peaks or dips does not only depend on the interaction force, but also depends on
 production amplitudes~\cite{Dong:2020hxe}. That is to say it is not obvious that these $P_c$ states in the molecular picture would favor their production in $\Lambda_b\to J/\psi p K$. Since $\Lambda_b$ is an isoscalar, the production of $\Sigma_{c}^{(*)}\bar{D}^{(*)}$ in the $\Lambda_b$ hadronic weak decays can only be through the light diquark violation process. This problem was firstly pointed out in Ref.~\cite{Liu:2015fea}. It was also mentioned in Ref.~\cite{Liu:2015fea} that, supposing that the $ud$ diquark within $\Lambda_b$ is a compact object, the production of these $P_c$ states will be extremely suppressed. In other words, the observation of these $P_c$ states in the hadronic molecule scenario implies that the $[ud]$ diquark cannot be a compact object in space. Instead, the quantum correlation between the light quarks should be the key for resolving the contradictions when putting all the dynamics together.

In order to further clarify the role played by the light diquark configuration in the production of the hidden charm $P_c$ states in the molecular picture, we investigate the two-body hadronic weak decays of the heavy flavor baryons and mesons, i.e. $\Lambda_b$, $\bar{B}_s^0$, $\Xi_b^0$, $\bar{B}^0$, $\Xi_b^-$ and $B^-$, by implementing the SU(3)-flavor symmetry. Actually, it has been pointed out that both the $\Lambda_c$ and the $\Sigma_c$ can be produced in $\Lambda_b$ decays~\cite{Liu:2015fea}.  Taking the emergence of hidden charm pentaquarks as an example,  the $\Lambda_{b}\to \Lambda_{c}^{(*)}\bar{D}^{(*)}K^{-}$ process can occur  with the $[ud]$ diquark as a spectator, while the $\Lambda_{b}\to \Sigma_{c}^{(*)}\bar{D}^{(*)}K^{-}$ process violates the diquark model and can happen via the two processes illustrated in the Fig.~1(b) and Fig.~1(c) of Ref.~\cite{Liu:2015fea}. This mechanism has also been pointed out by Ref.~\cite{Burns:2022uiv}. However, due to the non-perturbative  behavior of QCD, it is difficult to calculate the amplitude quantitatively.  In this paper, we investigate the validity of diquark model for the two-body decay of bottomed mesons and baryons,  which gives a hint of the significance for the $\Lambda_b\to \Sigma_{c}^{(*)}\bar{D}^{(*)}K^{-}$ process.  

The paper is organized as follows. In Sec.~\ref{Formalism} the framework is presented. Numerical results and related discussions are presented in Sec.~\ref{Results and discussions}. Sec.~\ref{Summary} is a brief summary. 

\section{Formalism}\label{Formalism}

\begin{figure}[h]
    \centering
    \subfigure[]{
    \includegraphics[width=0.3\textwidth]{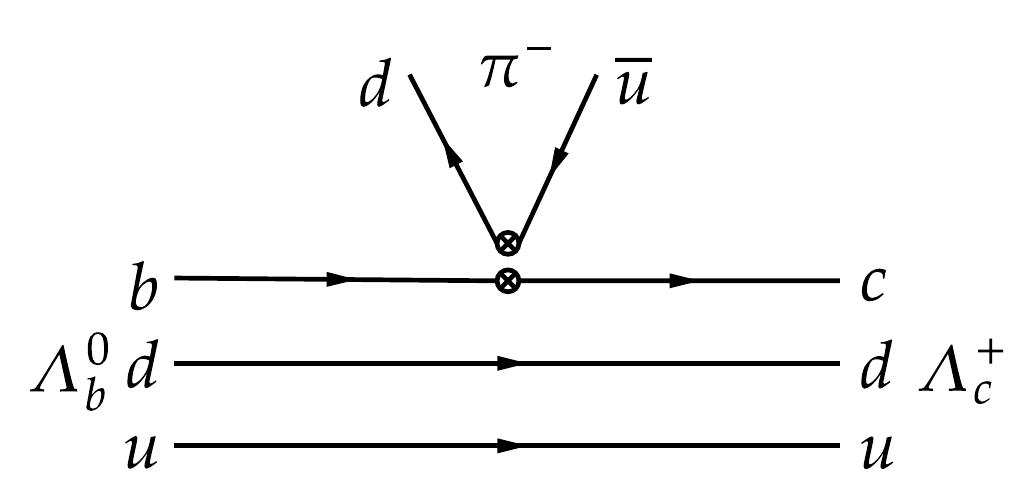}
    \label{fig1(a)}
    }
    \subfigure[]{
    \includegraphics[width=0.3\textwidth]{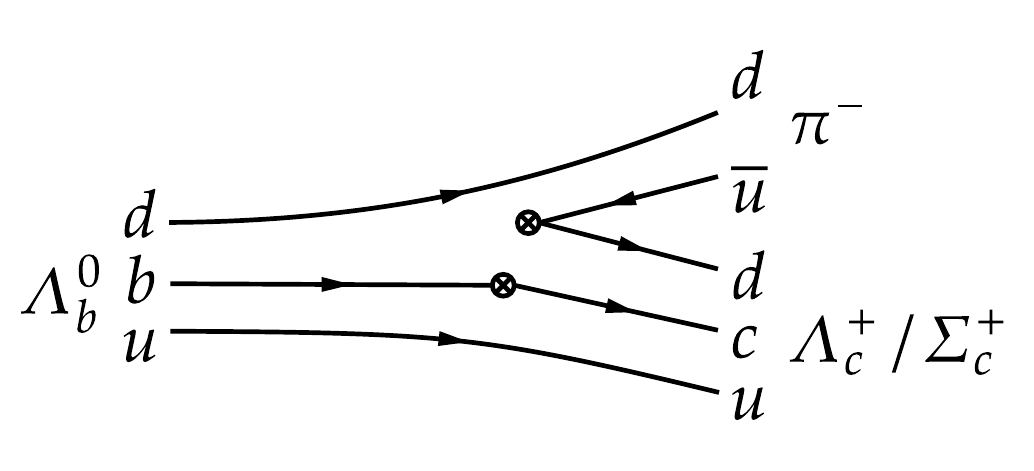}
    \label{fig1(b)}
    }
   \caption{The Feynman diagrams correspond to the decays $\Lambda_{b}^{0}\rightarrow \Lambda_{c}^{+}\pi^{-}$ and $\Lambda_{b}^{0}\rightarrow \Sigma_{c}^{+}\pi^{-}$. The
   subfigures (a) and (b) are the diquark conserved and violation diagrams, respectively. The diagrams for $\Xi_b^0$ and $\Xi_b^-$ decays are analogous. The crossed point is the effective weak decay vertex.}
   \label{fig1}
\end{figure}

  For the lack of the experiment data of three-body decay for bottomed hadrons in Review of Particle Physics ~\cite{ParticleDataGroup:2022pth},
  we resort to their two-body decays. Since the spin singlet $[ud]$, $[us]$, $[sd]$ diquarks conjugate to $\bar{s}$, $\bar{d}$, $\bar{u}$ quarks, respectively, according to the SU(3) group theory,
  one can simultaneously study the decays of $\Lambda_b^{0}$, $\Xi_b^0$, $\Xi_b^-$,
  $\bar{B}_{s}^{0}$, $\bar{B}^0$ and $B^-$. Their life time, total decay widths and masses are collected in App.~\ref{AppendixA}. In order to construct a simple but more general model, we only consider the contributions from the dominant tree-level diagrams, while those from the $W$-exchanged, annihilation or penguin diagrams are usually expected to be small \cite{Leibovich:2003tw,Fan:2012kn,Zhang:2022iun} and have been neglected in our calculations. For instance, the decays $\bar{B}_s^0 \to D^0\pi^0$, $D^+D^-$, etc. which can occur only through the annihilation diagrams, won' t be analyzed in the current work.
The two-body decay partial width reads
\begin{eqnarray}
\mathrm{d}\Gamma = \frac{(2\pi)^{4}}{2M}\left|\mathcal{\bar{M}}\right|^{2}\mathrm{d}\Phi_2(p_1;p_2, p_3),
\label{eq:width}
\end{eqnarray}
with $M$ the mass of decaying particle, $\mathcal{\bar{M}}$ the spin averaged decay amplitude, $\mathrm{d}\Phi_2$ the two-body phase space. 
Here $p_1$, $p_2$ and $p_3$ are the momenta of the initial decay particle
and the two final ones, respectively. As the polarization of the initial
particle and final particles will be averaged and summed, 
the phase space integration is trivial and gives
\begin{eqnarray}
    \Gamma = \frac{1}{8\pi M^{2}}\left|\vec{p}\right|\left|\mathcal{\bar{M}
    }\right|^{2}.
    \label{eq:two-body decay width}
\end{eqnarray}
 Here, $\vec{p}$ is the momentum of the two final states in the rest
 frame of decaying particle. For the decays whose partial decay widths are not measured in experiment, their momenta can be obtained by
\begin{eqnarray}
|\vec{p}|=\frac{\sqrt{[M^{2}-(m_{1}+m_{2})^{2}][M^{2}-(m_{1}-m_{2})^{2}]}}{2M},
\end{eqnarray}
with $m_{1,2}$
the masses corresponding to the two final particles, respectively. The decay amplitude $\bar{\mathcal{M}},$
 will be parameterized as the diquark conserved and violation 
 parts, for which we use the $\Lambda_b$ and $\bar{B}_s^0$ decays as an example to illustrate the parameterization scheme. The decay mechanism is based on the leading order contribution of the effective weak Hamiltonian~\cite{Buchalla:1995vs}, i.e. the $W$-emitted diagram.
 The diquark conserved and violation diagrams
 are presented in Fig.~\ref{fig1}. The diquark conserved
 amplitude shown in Fig.~\ref{fig1(a)} reads
 \begin{eqnarray}
     \bar{\mathcal{M}}^\mathrm{conversed}(\Lambda_b^0\to \Lambda_c^+\pi^-)=\frac{G_F}{\sqrt{2}}V_{cb}V_{ud}^*\mathcal{M}_a ,
 \end{eqnarray}
 with Fermi coupling constant $G_F=1.17\times 10^{-5}$ GeV$^{-2}$~\cite{ParticleDataGroup:2022pth} 
 and the factor $\frac{1}{\sqrt{2}}$ from the standard effective weak Hamiltonian~\cite{Li:2008ts}, and 
 $\mathcal{M}_a$ the amplitude parameter for the diagram respecting the light diquark. 
 As shown by Fig.~\ref{fig1(a)}, the $[ud]$ diquark behaves as a spectator.
 $V_{cb}$ and $V^{*}_{ud}$ are the Cabibbo-Kobayashi-Maskawa (CKM) matrix elements. 
  The diquark violation diagram is shown by Fig.~\ref{fig1(b)}
 and the contribution of its decay to $\Lambda_c^+\pi^-$ and $\Sigma_c^+\pi^-$ can be obtained by Rach rearrangement, which has been successfully used for double charm baryon decays~\cite{Wang:2017mqp, Ke:2019lcf}. In Fig.~\ref{fig1(b)}, both the isospin and spin of the initial $[ud]$ light diquark in $\Lambda_b$ are zero,
 which is denoted as $|ud_1\rangle_{(0,0)}$. The two subindexes in the bracket are for isospin and its third component. The $|d_2\bar{u}\rangle_{(1,-1)}$ diquark
 with isospin $1$ and its third component $-1$ is created by 
 the weak effective vertex or more fundamentally by the $W$-emitted diagram.
 The subindexes $1$ and $2$ of $d$ quark mean that the $d$ quark is from 
 the initial bottom baryon and created from weak vertex, respectively.
 By performing Rach rearrangement(for details see App.~\ref{AppendixB}), one obtains
 \begin{widetext}
  \begin{equation}
    \begin{split}
      &{\left|u{d}_{1}\right\rangle}_{(0, 0)}{\left|{d}_{2}\bar{u}\right\rangle}_{(1, -1)}\\
      =&\frac{1}{2}{\left|u{d}_{2}\right\rangle}_{(0, 0)}{\left|{d}_{1}\bar{u}\right\rangle}_{(1, -1)}-\frac{1}{2}{\left|u{d}_{2}\right\rangle}_{(1, -1)}{\left|{d}_{1}\bar{u}\right\rangle}_{(0, 0)}+\frac{1}{\sqrt{2}}{\left|u{d}_{2}\right\rangle}_{(1, 0)}{\left|{d}_{1}\bar{u}\right\rangle}_{(1, -1)}\\
      \Rightarrow&\frac{1}{2}\Lambda_{c}^{+}\pi^{-}-\frac{1}{2}\Sigma_{c}^{0}\eta+\frac{1}{\sqrt{2}}\Sigma_{c}^{+}\pi^{-}.
    \end{split}
    \label{eq:Rach}
  \end{equation}
  \end{widetext}
  Note that the isospins of the $[ud_2]$ diquark in the first and third terms are singlet and triplet, respectively, and will be hadronized into $\Lambda_c^+$ and $\Sigma_c^+$.
  Following Eq.~\eqref{eq:Rach}, one can obtain the decay amplitudes
\begin{eqnarray}
    \bar{\mathcal{M}}^\mathrm{violation}(\Lambda_b\to \Lambda_c^+\pi^-)& = &\frac{G_F}{\sqrt{2}}V_{cb}V_{ud}^*\frac{1}{2}\mathcal{M}_b, \\
    \bar{\mathcal{M}}^\mathrm{violation}(\Lambda_b\to \Sigma_c^0\eta)& = &-\frac{G_F}{\sqrt{2}}V_{cb}V_{ud}^*\frac{1}{2}\mathcal{M}_b, \\
    \bar{\mathcal{M}}^\mathrm{violation}(\Lambda_b\to \Sigma_c^+\pi^-)& = &\frac{G_F}{\sqrt{2}}V_{cb}V_{ud}^*\frac{1}{\sqrt{2}}\mathcal{M}_b, 
\end{eqnarray}
  from the light diquark violation diagram, with 
  $\mathcal{M}_b$ the amplitude parameter for the diagram breaking the light diquark. 
The total amplitude for a given process is the sum of the 
light diquark conserved and violation contributions, e.g.
\begin{equation}
\begin{split}
    &\bar{\mathcal{M}}(\Lambda_b^0\to\Lambda_c^+\pi^-)\\
    =&\bar{\mathcal{M}}^\mathrm{conserved}(\Lambda_b\to \Lambda_c^+\pi^-)+\bar{\mathcal{M}}^\mathrm{violation}(\Lambda_b\to \Lambda_c^+\pi^-)\\
    =&\frac{G_F}{\sqrt{2}}V_{cb}V_{ud}^*\left(\mathcal{M}_a+\frac{1}{2}\mathcal{M}_b\right). 
\end{split}
\end{equation}
In general, $\mathcal{M}_a$ and $\mathcal{M}_b$ are complex numbers, but the physical observables are the modulo square of the above equation, so it is trivial to notice that there are actually only 3 degrees of freedom, i.e. the physical observables depend only on the relative phase between $\mathcal{M}_a$ and  $\mathcal{M}_b$. Therefore, the decay amplitude can be further parameterized as
\begin{eqnarray}
    \mathcal{M}_a=a,\quad \mathcal{M}_b=c+id,
\end{eqnarray}
with $a$, $c$, $d$ real fit parameters, the relative strength between the imaginary part $d$ and the real part $c$ describe the relative phase. 

\begin{figure}[h]
    \centering
    \subfigure[]{
    \includegraphics[width=0.3\textwidth]{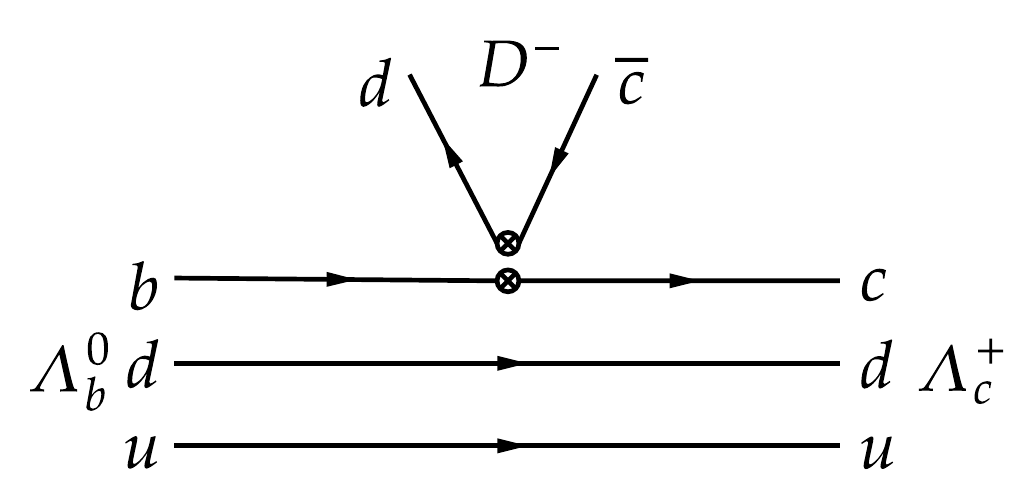}
    \label{fig2(a)}
    }
    \subfigure[]{
    \includegraphics[width=0.3\textwidth]{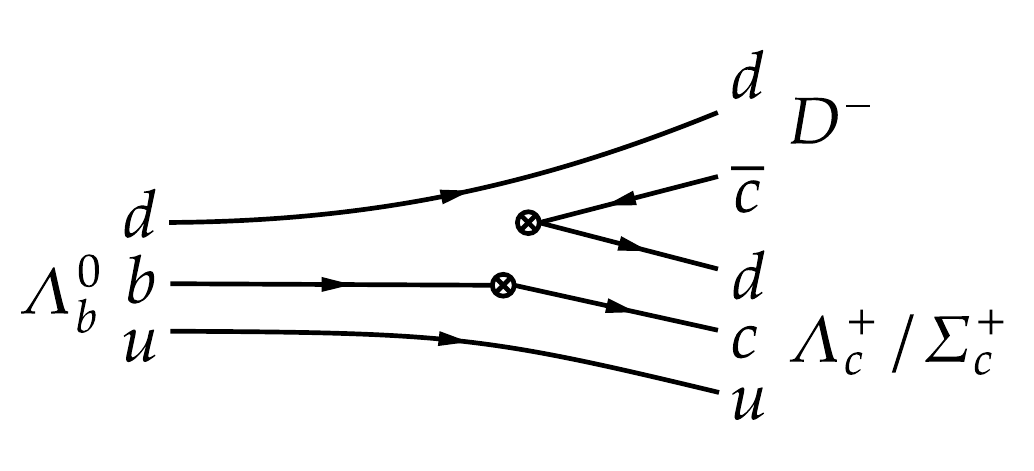}
    \label{fig2(b)}
    }
   \caption{The Feynman diagrams correspond to the decays $\Lambda_{b}^{0}\rightarrow \Lambda_{c}^{+}D^{-}$ and $\Lambda_{b}^{0}\rightarrow \Sigma_{c}^{+}D^{-}$. The
   subfigures (a) and (b) are the diquark conserved and violation diagrams, respectively. The diagrams for $\Xi_b^0$ and $\Xi_b^-$ decays are analogous. The crossed point is the effective weak decay vertex.}
   \label{fig2}
\end{figure}
\begin{figure}
    \centering
    \subfigure[]{
    \includegraphics[width=0.3\textwidth]{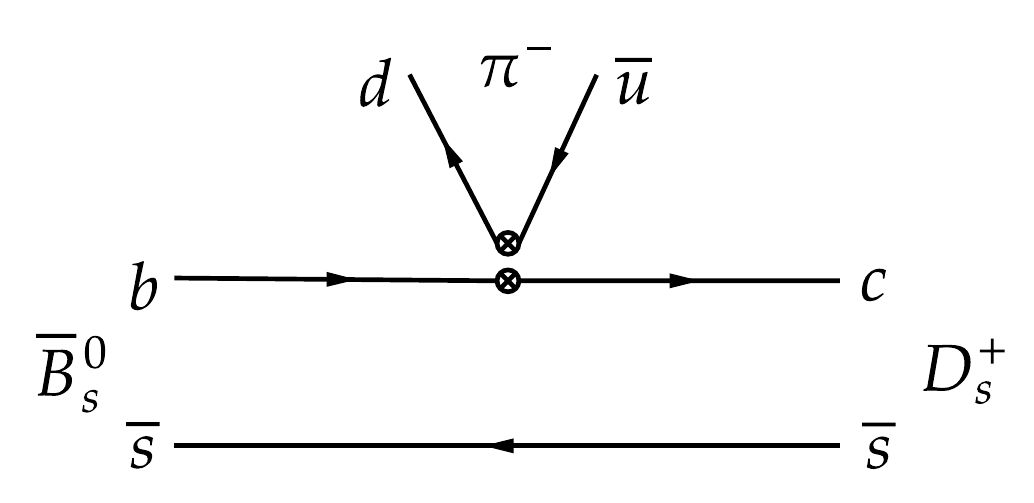}
    \label{fig3(a)}
    }
    \subfigure[]{
    \includegraphics[width=0.3\textwidth]{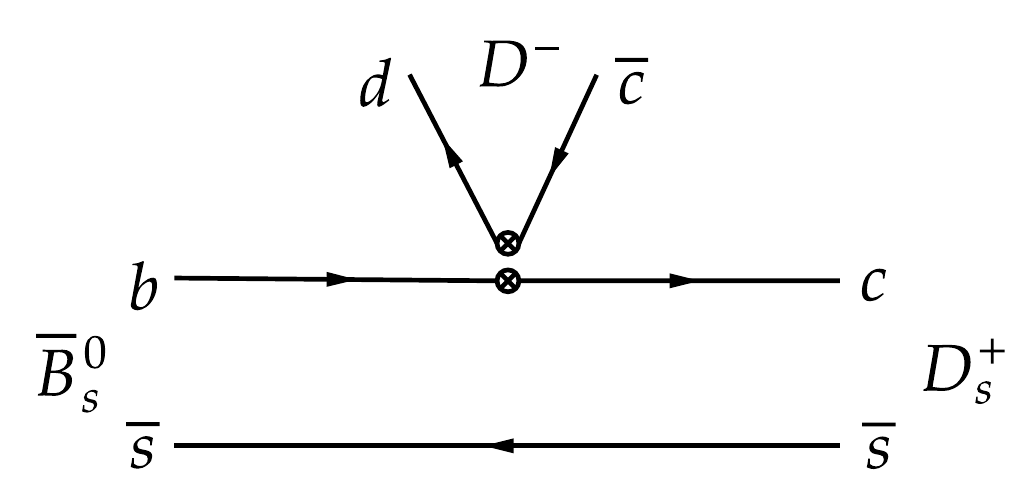}
    \label{fig3(b)}
    }   
    \caption{The Feynman diagrams correspond to decays $\bar{B}_{s}^{0}\rightarrow D_{s}^{+}\pi^{-}$(a) and $\bar{B}_{s}^{0}\rightarrow D_{s}^{+}D^{-}$(b).The diagrams for $\bar{B}^0$ and $B^-$ decays are analogous. The crossed point is the effective weak decay vertex. }
    \label{fig3}
\end{figure}
\begin{figure}
    \centering
    \subfigure[]{   \includegraphics[width=0.3\textwidth]{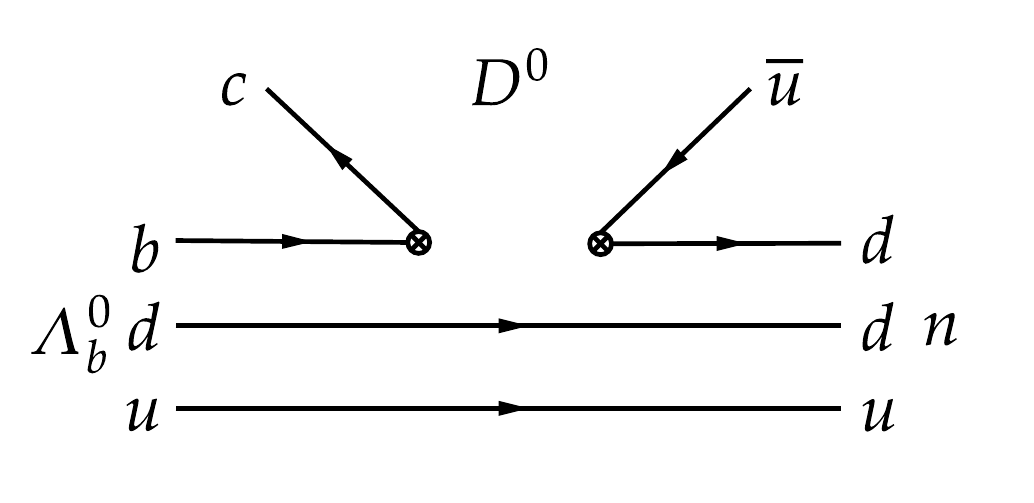}
    \label{fig4(a)}
    }
    \subfigure[]{
    \includegraphics[width=0.3\textwidth]{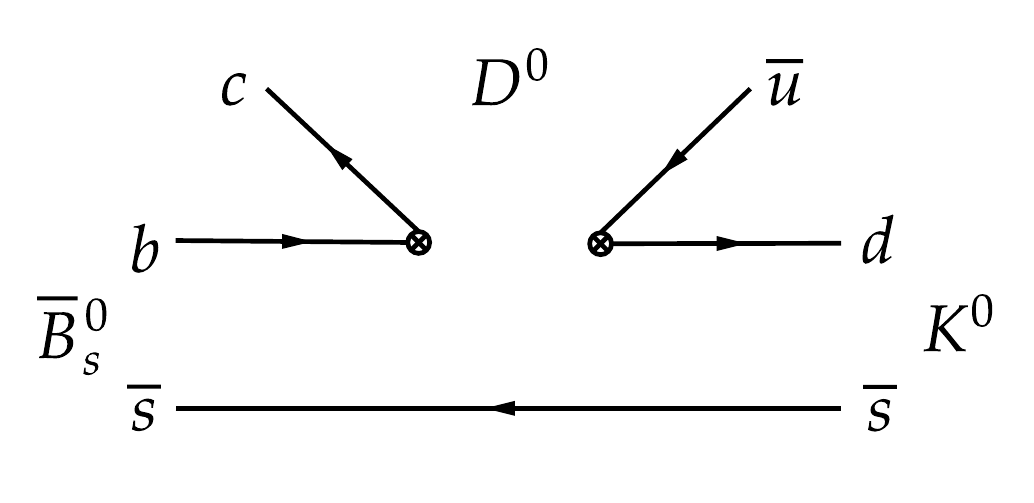}
    \label{fig4(b)}
    } 
    \caption{The Feynman diagrams correspond to the color-suppressed processes $\Lambda_{b}^{0}\rightarrow n D^{0}$ and $\bar{B}_{s}^{0}\rightarrow K^{0}D^{0}.$}
    \label{fig4}
\end{figure}
\begin{figure}[H]
\centering
\subfigure[]{
\includegraphics[width=0.3\textwidth]{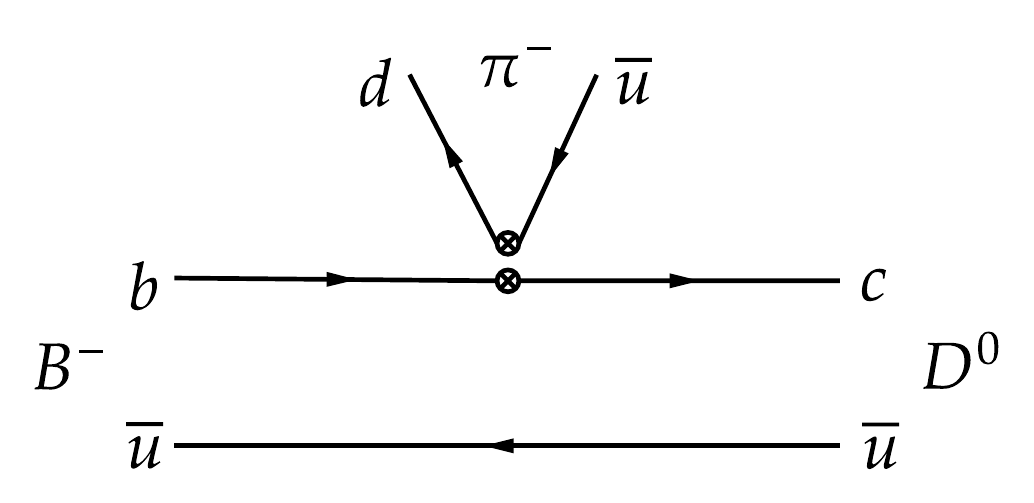}
}
\subfigure[]{
\includegraphics[width=0.3\textwidth]{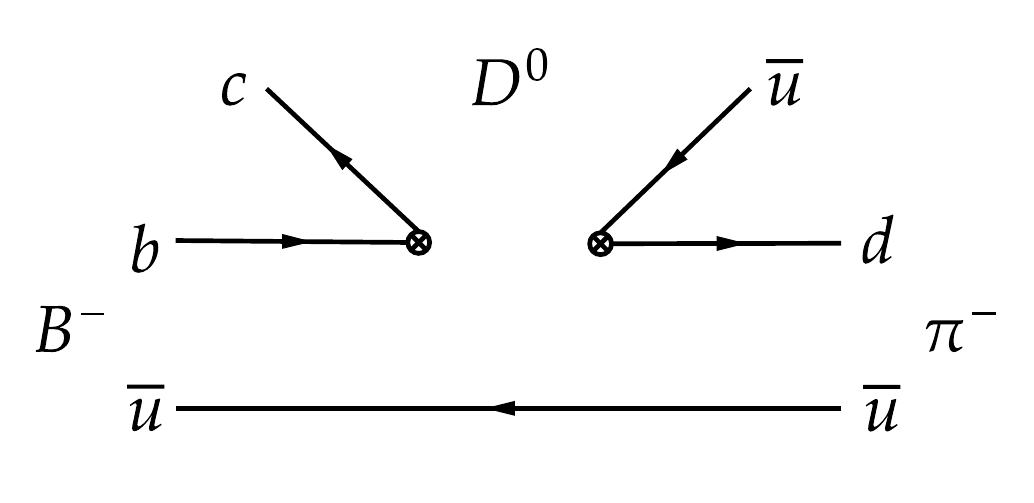}
}
\caption{The Feynman diagrams correspond to decay $B^{-}\rightarrow D^0\pi^{-}$. The subfigures (a) and (b) are the color-favored and color-suppressed processes, respectively. Replace the $d$ quark by $s$ quark then one can get the similar diagrams for $B^{-}\rightarrow D^0 K^{-}$.}
\label{fig5}
\end{figure}

Analogously, one can obtain the decay amplitude of the $\Lambda_b^0\to \Lambda_c^+ D^-$ process
as shown by Fig.~\ref{fig2}.
Fig.~\ref{fig2(a)} respects light diquark and gives the amplitude 
\begin{eqnarray}
    \bar{\mathcal{M}}^\mathrm{conserved}(\Lambda_b^0\to\Lambda_c^+ D^-)=\frac{G_F}{\sqrt{2}}V_{cb}V_{cd}^* \mathcal{M}_a. 
\end{eqnarray}
Fig.~\ref{fig2(b)} breaks light diquark and gives the amplitude 
\begin{eqnarray}
    \bar{\mathcal{M}}^\mathrm{violation}(\Lambda_b^0\to\Lambda_c^+ D^-)=-\frac{1}{2}\frac{G_F}{\sqrt{2}}V_{cb}V_{cd}^* \mathcal{M}_b.
\end{eqnarray}
The prefactor '$-\frac{1}{2}$' is from the recoupling 
 \begin{equation}
    \begin{split}
      &{\left|u{d}_{1}\right\rangle}_{(0, 0)}{\left|{d}_{2}\right\rangle}_{(\frac{1}{2}, -\frac{1}{2})}\\
      =&-\frac{1}{2}{\left|u{d}_{2}\right\rangle}_{(0, 0)}{\left|{d}_{1}\right\rangle}_{(\frac{1}{2}, -\frac{1}{2})}+\frac{\sqrt{3}}{2}{\left|u{d}_{2}\right\rangle}_{(1, 1)}{\left|{d}_{1}\right\rangle}_{(\frac{1}{2}, -\frac{1}{2})}\\
      \Rightarrow&-\frac{1}{2}\Lambda_{c}^{+}D^{-}+\frac{\sqrt{3}}{2}\Sigma_{c}^{+}D^{-}.
    \end{split}
    \label{eq:3q}
  \end{equation}
  The second $d$ quark in $|d_2\rangle_{(\frac{1}{2},-\frac{1}{2})}$
   is from the emitted $W$ boson as shown by Fig.~\ref{fig2(b)}. Here the $6-j$ symbol for recoupling of three quarks is used and the details can be found in App.~\ref{AppendixB}.
If the diquark picture works well, Fig.~\ref{fig1(a)}
and Fig.~\ref{fig2(a)} correspond to the $\bar{B}_s^0\to D_s^+\pi^-$ and $\bar{B}_s^0\to D_s^+D^-$ processes as shown by Fig.~\ref{fig3}.
The decay $\bar{B}_s^0\rightarrow K^{0}D^{0}$ corresponds to the $\Lambda_{b}^{0}\rightarrow nD^{0}$ in the light diquark picture as shown by Fig.~\ref{fig4} and is a color-suppressed process, 
as the quark pair emitted from $W$ boson goes into different hadrons. 
As a result, another parameter $F$ is considered to take into account the color-suppressed effect and the amplitude reads
\begin{equation}
\bar{\mathcal{M}}=\frac{G_F}{\sqrt{2}}V_{cb}V_{ud}^{*}F \mathcal{M}_{a}.
\end{equation}
Specially, there are both color-favored and color-suppressed processes for the decays $B^-\rightarrow D^0\pi^-$ and $B^-\rightarrow D^0 K^-$ presented in Fig.~\ref{fig5}, so their amplitudes are written as 
\begin{eqnarray}
    \bar{\mathcal{M}}(B^-\rightarrow D^0\pi^-)&=&\frac{G_F}{\sqrt{2}}V_{cb}V_{ud}^{*}(1+F)\mathcal{M}_{a},\\
    \bar{\mathcal{M}}(B^-\rightarrow D^0 K^-)&=&\frac{G_F}{\sqrt{2}}V_{cb}V_{us}^{*}(1+F)\mathcal{M}_{a}.
\end{eqnarray}
The decay amplitudes for other related processes can be found in Tab.~\ref{tab1:amplitude expressions}. It is worth noting that the decay constant ratios $R_1 = f_K/f_{\pi}=1.20$, $R_2=f_D/f_{\pi}=1.56$ and $R_3 =f_{D_s}/f_{\pi}=1.98$~\cite{Chang:2018aut} are multiplied to the corresponding channels to consider the effect of different final states on the diquark-conserved part amplitude $\mathcal{M}_a$, unless that the decay channel is color-suppressed. This factor also takes into account the SU(3) flavor symmetry breaking effect discussed in Refs.~\cite{Qin:2013tje,Jia:2019zxi,Huber:2021cgk}. With these decay amplitudes, one can obtain the corresponding decay widths via Eq.~\eqref{eq:two-body decay width}. 
One should notice that our frame work is different from the pQCD calculation~\cite{Li:2008ts} of the bottomed meson decays with explicitly considering the distribution amplitudes of the initial and final mesons, and the contribution of the other topological diagrams. In this work,
we aim at testing to which extent the light diquark picture works for heavy baryon decays. Thus, we only consider the leading order tree diagram contribution, but without the distribution amplitudes of the initial and final hadrons. 
\begin{center}
\begin{table}[h]
    \centering
    \caption{The decay amplitudes of bottomed baryons and bottomed mesons.
    }
    \label{tab1:amplitude expressions}
    \begin{tabular}{l|c}
    \hline\hline
 \textbf{Decay mode}  &  \textbf{Expression of $\bar{\mathcal{M}}$}  \\
    \hline
    \hline
   $\Lambda_b^0\rightarrow \Lambda_c^{+}\pi^{-}$  & $\frac{G_{F}}{\sqrt{2}}V_{cb}V_{ud}^{*}(\mathcal{M}_{a}+\frac{1}{2}\mathcal{M}_{b})$\\
   $\Lambda_b^0\rightarrow \Sigma_c^{+}\pi^{-}$  & $\frac{G_{F}}{\sqrt{2}}V_{cb}V_{ud}^{*}\frac{1}{\sqrt{2}}\mathcal{M}_{b}$\\
   $\Lambda_b^0\rightarrow nD^{0}$  &  $\frac{G_{F}}{\sqrt{2}}V_{cb}V_{ud}^{*}F\mathcal{M}_{a}$ \\
   $\Lambda_b^0\rightarrow \Lambda_c^{+}D^{-}$  & $\frac{G_{F}}{\sqrt{2}}V_{cb}V_{cd}^{*}(R_2\mathcal{M}_{a}-\frac{1}{2}\mathcal{M}_{b})$\\
   $\Lambda_b^0\rightarrow \Sigma_c^{+}D^{-}$  & $\frac{G_{F}}{\sqrt{2}}V_{cb}V_{cd}^{*}\frac{\sqrt{3}}{2}\mathcal{M}_{b}$\\ 
   $\bar{B}_s^0\rightarrow D^{+}_{s}\pi^{-}$   &  $\frac{G_{F}}{\sqrt{2}}V_{cb}V_{ud}^{*}\mathcal{M}_{a}$\\
   $\bar{B}_s^0\rightarrow D^{0}K^{0}$   &  $\frac{G_{F}}{\sqrt{2}}V_{cb}V_{ud}^{*}F\mathcal{M}_{a}$\\
   $\bar{B}_s^0\rightarrow D^{+}_{s}D^{-}$   &  $\frac{G_{F}}{\sqrt{2}}V_{cb}V_{cd}^{*}R_2\mathcal{M}_{a}$\\
   \hline 
   $\Lambda_b^0\rightarrow \Lambda_c^{+}K^{-}$  & $\frac{G_{F}}{\sqrt{2}}V_{cb}V_{us}^{*}R_1\mathcal{M}_{a}$\\
   $\Lambda_b^0\rightarrow \Lambda_c^{+}D_{s}^{-}$  & $\frac{G_{F}}{\sqrt{2}}V_{cb}V_{cs}^{*}R_3\mathcal{M}_{a}$\\
   $\bar{B}_s^0\rightarrow D_{s}^{+}K^{-}$   &  $\frac{G_{F}}{\sqrt{2}}V_{cb}V_{us}^{*}R_1\mathcal{M}_{a}$\\
   $\bar{B}_s^0\rightarrow D_{s}^{+}D_{s}^{-}$   &  $\frac{G_{F}}{\sqrt{2}}V_{cb}V_{cs}^{*}R_3\mathcal{M}_{a}$\\\hline
   $\Xi_{b}^0\rightarrow \Xi_{c}^{+}\pi^{-}$   &  $\frac{G_{F}}{\sqrt{2}}V_{cb}V_{ud}^{*}\mathcal{M}_{a}$\\
   $\Xi_{b}^0\rightarrow \Xi_{c}^{+}K^{-}$   &  $\frac{G_{F}}{\sqrt{2}}V_{cb}V_{us}^{*}R_1\mathcal{M}_{a}$\\
   $\Xi_{b}^0\rightarrow \Xi_{c}^{+}D^{-}$   &  $\frac{G_{F}}{\sqrt{2}}V_{cb}V_{cd}^{*}R_2\mathcal{M}_{a}$\\
   $\Xi_{b}^0\rightarrow \Xi_{c}^{+}D_{s}^{-}$   &  $\frac{G_{F}}{\sqrt{2}}V_{cb}V_{cs}^{*}R_3\mathcal{M}_{a}$\\
   $\bar{B}^0\rightarrow D^{+}\pi^{-}$   &  $\frac{G_{F}}{\sqrt{2}}V_{cb}V_{ud}^{*}\mathcal{M}_{a}$\\
   $\bar{B}^0\rightarrow D^{+}K^{-}$   &  $\frac{G_{F}}{\sqrt{2}}V_{cb}V_{us}^{*}R_1\mathcal{M}_{a}$\\
   $\bar{B}^0\rightarrow D^{+}D^{-}$   &  $\frac{G_{F}}{\sqrt{2}}V_{cb}V_{cd}^{*}R_2\mathcal{M}_{a}$\\
   $\bar{B}^0\rightarrow D^{+}D_{s}^{-}$   &  $\frac{G_{F}}{\sqrt{2}}V_{cb}V_{cs}^{*}R_3\mathcal{M}_{a}$\\\hline
   $\Xi_{b}^{-}\rightarrow \Xi_{c}^{0}\pi^{-}$   &  $\frac{G_{F}}{\sqrt{2}}V_{cb}V_{ud}^{*}(\mathcal{M}_{a}+\mathcal{M}_{b})$\\
   $\Xi_{b}^{-}\rightarrow \Xi_{c}^{0}K^{-}$   &  $\frac{G_{F}}{\sqrt{2}}V_{cb}V_{us}^{*}R_1\mathcal{M}_{a}$\\
   $\Xi_{b}^{-}\rightarrow \Xi_{c}^{0}D^{-}$   &  $\frac{G_{F}}{\sqrt{2}}V_{cb}V_{cd}^{*}(R_2\mathcal{M}_{a}+\mathcal{M}_{b})$\\
   $\Xi_{b}^{-}\rightarrow \Xi_{c}^{0}D_{s}^{-}$   &  $\frac{G_{F}}{\sqrt{2}}V_{cb}V_{cs}^{*}R_3\mathcal{M}_{a}$\\
   $B^{-}\rightarrow D^{0}\pi^{-}$   &  $\frac{G_{F}}{\sqrt{2}}V_{cb}V_{ud}^{*}(1+F)\mathcal{M}_{a}$\\
   $B^{-}\rightarrow D^{0}K^{-}$   &  $\frac{G_{F}}{\sqrt{2}}V_{cb}V_{us}^{*}(R_1+F)\mathcal{M}_{a}$\\
   $B^{-}\rightarrow D^{0}D^{-}$   &  $\frac{G_{F}}{\sqrt{2}}V_{cb}V_{cd}^{*}R_2\mathcal{M}_{a}$\\
   $B^{-}\rightarrow D^{0}D_{s}^{-}$   &  $\frac{G_{F}}{\sqrt{2}}V_{cb}V_{cs}^{*}R_3\mathcal{M}_{a}$\\   
   \hline
   \end{tabular}     
\end{table}
\end{center} 

\section{Results and discussions} \label{Results and discussions}
After fitting into the experimental decay widths of $\Lambda_{b}^{0}\to\Lambda_c^{+}\pi^{-}/\Lambda_{c}^{+}D^{-}/\Lambda_{c}^{+}D_{s}^{-}$ and $\bar{B}_{s}^{0}\to D_{s}^{+}\pi^{-}/D_{s}^{+}D^{-}/D^{0}K^{0}$, we obtain our fit results (as shown by Fig.~\ref{fig3:fitting results}) and the values of the parameters (as shown in Tab.~\ref{tab2:parameter}). 
As shown in Fig.~\ref{fig3:fitting results}, our model describes the experimental data with $\chi^{2} / \mathrm{d.o.f.}=1.2$. The fitted data matches well with the experimental data, and the reason why $\chi^{2} / \mathrm{d.o.f.}$ is not very small is that the accuracy of the experimental measurements varies considerably from one channel to another, and our model ignores some higher-order corrections for the different channels.
The value of the dimensionless parameter $F$ is $0.372\pm0.039$, whose absolute value is approximate to the factor $\frac{1}{3}$ reflecting the color-suppressed contribution. Since $\left|\mathcal{M}_{a}\right|:\left|\mathcal{M}_{b}\right|=1.973:\sqrt{(-0.63)^{2}+4.09^{2}}\approx 0.476$, the diquark model is greatly violated in these heavy hadron decays, which supports that the light diquark violated channel, i.e. $\Sigma_c^{(*)}\bar{D}^{(*)}$ channels, could be produced in $\Lambda_b^0$ decays with a sizable amplitude. 
The corresponding branching ratio between the $\Sigma_c^+D^-$ and the $\Lambda_c^+D^-$ channels
\begin{eqnarray}
    \frac{\mathcal{B}r(\Lambda_b^0\to \Sigma_cD^-)}{\mathcal{B}r(\Lambda_b^0\to \Lambda_cD^-)}=0.78
\end{eqnarray}
 also indicates that the peak structures in $\Lambda_{b}\rightarrow \Sigma_{c}^{*}\bar{D}^{(*)}K^{-}\rightarrow J/\Psi p K^{-}$ could be viewed as the $\Sigma_c^{(*)}\bar{D}^{(*)}$ hadronic molecular candidates. 
Several realistic fits~\cite{Du:2019pij,Du:2021fmf,Kuang:2020bnk,Nakamura:2021dix}
to the $J/\psi p$ invariant mass distribution and a recent lattice calculation~\cite{Xing:2022ijm} also demonstrate that the $\Sigma_c^{(*)}\bar{D}^{(*)}$ channels are the dominant 
channels for the formation of hidden charm pentaquarks. 

With the fitted parameters, one can predict the decay widths of the diquark-violated decays, i.e. $\Lambda_b^0 \rightarrow \Sigma_{c}^{+}\pi^{-}$ and $\Lambda_b^0 \rightarrow \Sigma_{c}^{+}D^{-}$ as shown in Tab.~\ref{tab3:prediction}. 
As the charged pion can be detected directly in the detector, 
the former one is more easier to be measured.  
The predicted width of the 
$\bar{B}_{s}^{0}\rightarrow D_{s}^{+}K^{-}$ process is in good agreement with the experimental data within the margin error, while the predicted width of $\bar{B}_{s}^{0}\rightarrow D^{+}_{s}D_{s}^{-}$ is much larger than the experimental data and that of $\Lambda_b^0\rightarrow\Lambda_c^+K^-$ is smaller than the experimental one, which are mainly due to the neglect of the other topological diagrams and the rough treatment of decay constant ratios between different decay channels in our model. Further experimental 
inputs, especially the three hadron decays, will give us a profound understanding of diquark model and the dynamics of the exotic hadrons.
Under SU(3) symmetry, the Cabibbo-allowed decay widths of the $\Xi_b^{0}$, $\Xi_b^{-}$, $\bar{B}^{0}$ and $B^{-}$ are also calculated, and are showed in Tab.~\ref{tab3:prediction}. Apart from that,
our calculations of the  $\bar{B}^{0}\to D^{+}\pi^{-}$ and $B^-\to D^0D^-$ decays are consistent well with the current experimental data within the errors,
which can be seen easily from the Tab.~\ref{tab3:prediction}. Some other decays that cannot be well described by our model are probably due to the significant effect of higher-order contributions such as annihilation, penguin, $W$-exchanged diagrams, the rough treatment of decay constant ratios between different decays and the effect of distribution amplitudes of the initial and final hadrons.
The errors in Tab.~\ref{tab3:prediction} are estimated according to Eq.~\eqref{eq:two-body decay width}, i.e. the total error of the decay width is from the uncertainties of the input masses and parameters. The error of momentum for the predicted decays can be calculated via 
$$\Delta |\vec{p}|=[(\frac{\partial{|\vec{p}|}}{\partial{M}}\cdot\Delta M)^{2}+(\frac{\partial{|\vec{p}|}}{\partial{m_{1}}}\cdot\Delta m_{1})^{2}+(\frac{\partial{|\vec{p}|}}{\partial{m_{2}}}\cdot\Delta m_{2})^{2}]^{\frac{1}{2}}.$$
with $M$, $m_{1}$ and $m_2$ the masses of initial and final hadrons. Since the errors from masses, momentum, and $G_{F}$ are very tiny, we ignore their contributions to the total errors of the final results. Thus, the error of each decay width $\Gamma_{i}$ reads
\begin{equation}
\begin{split}
\Delta \Gamma_{i} =&[(\frac{\partial{\Gamma_{i}}}{\partial{V_{cb}}}\cdot\Delta V_{cb})^{2}+(\frac{\partial{\Gamma_{i}}}{\partial{V_{qq'}}}\cdot\Delta V_{qq'})^{2}+(\frac{\partial{\Gamma_{i}}}{\partial{a}}\cdot\Delta a)^{2}\\
&+(\frac{\partial{\Gamma_{i}}}{\partial{c}}\cdot\Delta c)^{2}+(\frac{\partial{\Gamma_{i}}}{\partial{d}}\cdot\Delta d)^{2}+(\frac{\partial{\Gamma_{i}}}{\partial{F}}\cdot\Delta F)^{2}]^\frac{1}{2}.
\end{split}
\end{equation} 
\begin{center}
\begin{table}[h]
    \centering
    \caption{The value of each parameter in our fit.}
    \label{tab2:parameter}
    \begin{tabular}{l|c}
    \hline\hline
   \textbf{Parameter}  &  \textbf{Value}  \\
    \hline
    \hline
   $a~$[$\mathrm{GeV}$$^{3}$]  & $1.973\pm0.04$\\
   $c~$[$\mathrm{GeV}$$^{3}$]  & $-0.63\pm0.42$\\
   $d~$[$\mathrm{GeV}$$^{3}$]  & $4.09\pm0.39$\\
   $F$~(dimensionless) & $0.372\pm0.039$\\
   \hline
   \end{tabular}   
\end{table}
\end{center}
\section{Summary} \label{Summary}
There are numerous exotic candidates observed in experiments 
since the observation of the $X(3872)$ in 2003. Their  invariant mass distributions
do not only depend on the underlying formation dynamics,
but also depend on the production mechanism~\cite{Dong:2020hxe}. Lots of them are observed
in bottomed hadron decays, for instance, the hidden charm pentaquarks
observed in $\Lambda_b$ decays, the $X(3872)$ observed in $B$ decays. 
As a result, the study of bottomed hadron decays would provide
a key ingredient for the property of exotic hadrons. 
In this paper, we study the two-hadron decays of bottomed baryons 
and bottomed mesons to probe to which extent the light diquark picture works. That will give a constraint on the property of the exotic candidates observed in bottomed hadron decays. Our results show that the diquark picture is greatly violated in these
 decays since the diquark violated amplitude is significantly larger than the diquark conserved one.  More specifically, several theoretical studies of the hidden pentaquarks have a controversy whether the hidden charm pentaquark dominant channels $\Sigma_c^{(*)}\bar{D}^{(*)}$ could be produced in $\Lambda_b^0$ decays with sizable production rate or not. It is very crucial for its lineshape, i.e. either peak structures or dip structures~\cite{Dong:2020hxe}. Our work indicates that the $\Lambda_b$ has a sizable probability to decay into $\Sigma_c$, which gives an evidence that the hidden charm pentaquarks could express themselves as peaks in $\Lambda_b$ decays.
 In addition, the decay widths of the 
 $\Lambda_{b}^{0} \to \Sigma_{c}^{+}\pi^{-}$ and $\Lambda_{b}^{0} \to \Sigma_{c}^{+}D^{-}$ processes are also predicted  
 which can be measured in the forthcoming experiments.
  We also find that the predicted width of the 
$\bar{B}_{s}^{0}\rightarrow D_{s}^{+}K^{-}$ process is in good agreement with the experimental data within the margin error, while the predicted width of $\bar{B}_{s}^{0}\rightarrow D^{+}_{s}D_{s}^{-}$ is much larger than the experimental data and that of $\Lambda_b^0\rightarrow\Lambda_c^+K^-$ is smaller than the experimental one probably due to the neglect of the other topological diagrams and the rough treatment of decay constant ratios between different decays in our model.
\begin{widetext}
\begin{center}
\centering
\begin{figure}[h]  
\includegraphics[width=0.7\textwidth]{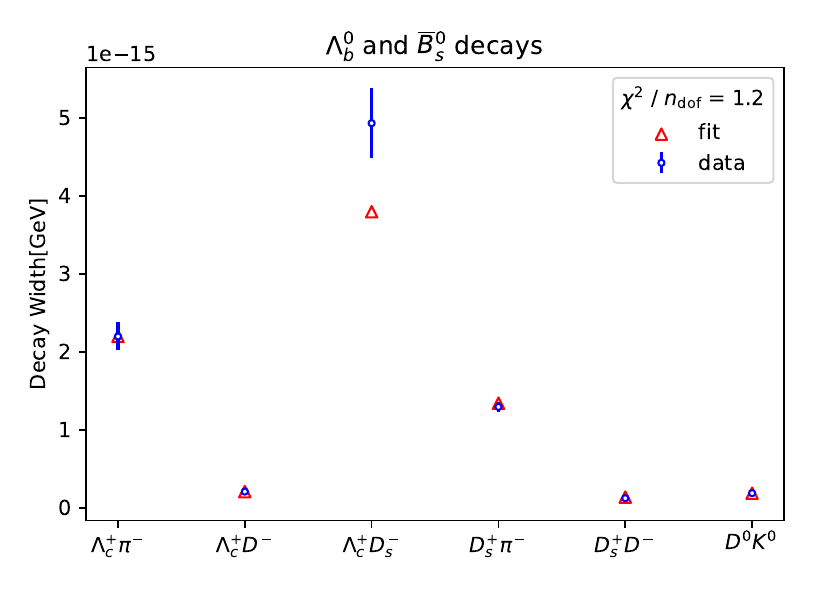}
\caption{The fitted decay widths with $\chi^{2} / \mathrm{d.o.f.}=1.2$ comparing with the experimental data. The blue circles are the experimental values. The red triangles are our results. }
\label{fig3:fitting results}
\end{figure}    
\end{center}   
\end{widetext}
{\bf Acknowledgements:}~~We are grateful to Qiang Zhao for the helpful discussion. 
This work is partly supported by 
the National Natural Science Foundation of China with Grant No.~12375073, No.~12035007, No.~12205106, No.~12105028, Guangdong Provincial funding with Grant No.~2019QN01X172,
Guangdong Major Project of Basic and Applied Basic Research No.~2020B0301030008,
 the NSFC and the Deutsche Forschungsgemeinschaft (DFG, German
Research Foundation) through the funds provided to the Sino-German Collaborative
Research Center TRR110 ``Symmetries and the Emergence of Structure in QCD"
(NSFC Grant No. 12070131001, DFG Project-ID 196253076-TRR 110). D.C.Y is also supported by the Natural Science Foundation of Jiangsu Province under Grant No. BK20200980.
 \begin{widetext}
\begin{center}
\begin{table}[ht]
    \centering
    \caption{The predicted widths, errors and corresponding experiment data of each decay mode(Units:GeV). These inputs correspond to the processes fitted in Fig.~\ref{fig3:fitting results}.}
    \label{tab3:prediction}
    \begin{tabular}{l|c|c|c}
    \hline
    \hline
\textbf{Decay mode}  &  \textbf{$\Gamma_{i}$(Theory)} &\textbf{Error} &\textbf{$\Gamma_{i}$(Experiment)}\\
    \hline
    \hline
$\Lambda_{b}^{0}\rightarrow\Lambda_{c}^{+}\pi^{-}$ &$-$  &$-$ &$(2.20\pm0.18)\times10^{-15}$\textbf{(Input)}\\
$\Lambda_{b}^{0}\rightarrow\Lambda_{c}^{+}D^{-}$ 
&$-$  &$-$  &$(2.06\pm0.27)\times10^{-16}$\textbf{(Input)}\\
$\bar{B}_{s}^{0}\rightarrow D_{s}^{+}\pi^{-}$ &$-$ &$-$ &$(1.30\pm0.06)\times10^{-15}$\textbf{(Input)}\\
$\bar{B}_{s}^{0}\rightarrow D^{0}K^{0}$ &$-$ &$-$ &$(1.86\pm0.39)\times10^{-16}$\textbf{(Input)}\\
$\bar{B}_{s}^{0}\rightarrow D_{s}^{+}D^{-}$ &$-$ &$-$ &$(1.21\pm0.22)\times10^{-16}$\textbf{(Input)}\\
$\Lambda_{b}^{0}\rightarrow\Sigma_{c}^{+}\pi^{-}$ &$2.63\times10^{-15}$ &$5.29\times10^{-16}$ &$-$\\
$\Lambda_{b}^{0}\rightarrow nD^{0}$ & $1.75\times10^{-16}$ & $3.93\times10^{-17}$ &$-$\\
$\Lambda_{b}^{0}\rightarrow\Sigma_{c}^{+}D^{-}$  
&$1.60\times10^{-16}$  &$3.26\times10^{-17}$ &$-$\\
\hline
$\Lambda_{b}^{0}\rightarrow\Lambda_{c}^{+}K^{-}$ 
&$9.31\times10^{-17}$  &$7.45\times10^{-18}$ &$(1.60\pm0.13)\times10^{-16}$\\
$\Lambda_{b}^{0}\rightarrow\Lambda_{c}^{+}D_{s}^{-}$ &$-$  &$-$  &$(4.93\pm0.45)\times10^{-15}$\textbf{(Input)}\\
$\bar{B}_{s}^{0}\rightarrow D_{s}^{+}K^{-}$ &$1.01\times10^{-16}$  &$8.09\times10^{-18}$  &$(9.76\pm0.52)\times10^{-17}$\\
$\bar{B}_{s}^{0}\rightarrow D_{s}^{+}D_{s}^{-}$ &$4.14\times10^{-15}$ 
&$3.34 \times10^{-16}$ &$(1.91\pm0.22)\times10^{-15}$\\
\hline
$\Xi_{b}^{0}\rightarrow\Xi_{c}^{+}\pi^{-}$  &$1.17\times10^{-15}$  
&$9.36\times10^{-17}$ &$-$\\ 
$\Xi_{b}^{0}\rightarrow\Xi_{c}^{+}K^{-}$    &$8.86\times10^{-17}$  
&$7.09\times10^{-18}$ &$-$\\
$\Xi_{b}^{0}\rightarrow\Xi_{c}^{+}D^{-}$     &$1.19\times10^{-16}$ 
&$1.04\times10^{-17}$ &$-$\\
$\Xi_{b}^{0}\rightarrow\Xi_{c}^{+}D_{s}^{-}$  &$3.61\times10^{-15}$
&$2.92\times10^{-16}$ &$-$\\
$\bar{B}^{0}\rightarrow D^{+}\pi^{-}$  &$1.38\times10^{-15}$ 
&$1.10\times10^{-16}$ &$(1.35\pm0.035)\times10^{-15}$\\
$\bar{B}^{0}\rightarrow D^{+}K^{-}$    &$1.04\times10^{-16}$ 
&$8.31\times10^{-18}$ &$(8.90\pm0.35)\times10^{-17}$\\
$\bar{B}^{0}\rightarrow D^{+}D^{-}$  &$1.39\times10^{-16}$ &$1.22\times10^{-17}$ &$(9.16\pm0.78)\times10^{-17}$\\
$\bar{B}^{0}\rightarrow D^{+}D_{s}^{-}$  &$4.25\times10^{-15}$ 
&$3.43\times10^{-16}$ &$(3.12\pm0.35)\times10^{-15}$\\
\hline
$\Xi_{b}^{-}\rightarrow\Xi_{c}^{0}\pi^{-}$  &$7.08\times10^{-15}$ 
&$1.26\times10^{-15}$ &$-$\\
$\Xi_{b}^{-}\rightarrow\Xi_{c}^{0}K^{-}$   &$8.85\times10^{-17}$  
&$7.09\times10^{-18}$ &$-$\\
$\Xi_{b}^{-}\rightarrow\Xi_{c}^{0}D^{-}$   &$3.81\times10^{-16}$ 
&$6.34\times10^{-17}$ &$-$\\
$\Xi_{b}^{-}\rightarrow\Xi_{c}^{0}D_{s}^{-}$  &$3.61\times10^{-15}$ 
&$2.91\times10^{-16}$ &$-$\\
$B^{-}\rightarrow D^{0}\pi^{-}$  
&$2.59\times10^{-15}$ 
&$2.54\times10^{-16}$
&$(1.88\pm0.05)\times10^{-15}$\\
$B^{-}\rightarrow D^{0}K^{-}$  
&$1.78\times10^{-16}$
&$1.68\times10^{-17}$
&$(1.48\pm0.06)\times10^{-16}$\\
$B^{-}\rightarrow D^{0}D^{-}$  
&$1.40\times10^{-16}$ 
&$1.22\times10^{-17}$
&$(1.53\pm0.16)\times10^{-16}$\\
$B^{-}\rightarrow D^{0}D_{s}^{-}$  
&$4.26\times10^{-15}$
&$3.43\times10^{-16}$
&$(3.62\pm0.36)\times10^{-15}$\\
    \hline
    \end{tabular}      
\end{table}
\end{center} 
\end{widetext}
\appendix
\section{Isospin Recoupling} \label{AppendixB}
The isospin recoupling of three particles reads
    \begin{equation}
    \begin{split}
     &\left|I_{1}(I_{2}I_{3})I_{23}Im_{I}\right\rangle\\
     =&\sum_{I_{12}}(-1)^{I_{1}+I_{2}+I+I_{3}}\sqrt{(2I_{12}+1)(2I_{23}+1)}\\
       &\cdot \begin{Bmatrix}
        I_{1} & I_{2} & I_{12}\\
        I_{3} & I     & I_{23}
        \end{Bmatrix}
        \left|(I_{1}I_{2})I_{12}I_{3}Im_{I}\right\rangle,   
    \end{split}
    \label{eq:6-j}
    \end{equation}
where $\left|I_{1}(I_{2}I_{3})I_{23}Im_{I}\right\rangle$ is the basis with $I_{i}$ the isospin of the $i$th quark, $I_{ij}$ the sum of isospin of the $i$th and the $j$th quark, $I$ the total isospin of three quarks and $m_{I}$ the third component of the total isospin. $\begin{Bmatrix}
        I_{1} & I_{2} & I_{12}\\
        I_{3} & I     & I_{23}
        \end{Bmatrix}$
is the $6-j$ symbol. With Eq.~\eqref{eq:6-j}, one can transform the basis $\left|I_{1}(I_{2}I_{3})I_{23}Im_{I}\right\rangle$ into the basis $\left|(I_{1}I_{2})I_{12}I_{3}Im_{I}\right\rangle$ and obtain Eq.~\eqref{eq:3q}.
\par
In addition, one can also obtain the isospin recoupling involving four quarks via
\begin{equation}
\begin{split}
&\left|(I_{1}I_{3})I_{13},(I_{2}I_{4})I_{24},Im_{I}\right\rangle\\
=& \sum_{I_{12},I_{34}}\sqrt{(2I_{12}+1)(2I_{34}+1)(2I_{13}+1)(2I_{24}+1)}\\
        &\cdot\begin{Bmatrix}
        I_{1} & I_{2} & I_{12}\\
        I_{3} & I_{4} & I_{34}\\
        I_{13}& I_{24}& I
        \end{Bmatrix}
        \left|(I_{1}I_{2})I_{12},(I_{3}I_{4})I_{34},Im_{I}\right\rangle.
\end{split}
\label{eq:9-j}
\end{equation}
Here $I$ is the total isospin of four quarks and $\begin{Bmatrix}
        I_{1} & I_{2} & I_{12}\\
        I_{3} & I_{4} & I_{34}\\
        I_{13}& I_{24}& I
        \end{Bmatrix}$ is the $9-j$ symbol. One can transform the basis $\left|(I_{1}I_{3})I_{13},(I_{2}I_{4})I_{24},Im_{I}\right\rangle$ into the basis $\left|(I_{1}I_{2})I_{12},(I_{3}I_{4})I_{34},Im_{I}\right\rangle$ and obtain Eq.~\eqref{eq:Rach}.
 \clearpage
 \section{Feynman Diagrams} \label{AppendixD}
\begin{figure}[H]
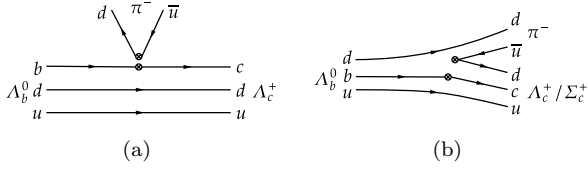

\centering
\subfigure[]{
\includegraphics[width=0.44\linewidth]{./image/fig1a.pdf}
}
\subfigure[]{
\includegraphics[width=0.44\linewidth]{./image/fig1b.pdf}
}
\caption{The Feynman diagrams of process $\Lambda_{b}^{0}\rightarrow\Lambda_{c}^{+}\pi^{-}$. The subfigure (a) is the diquark conserved diagram while (b) is the diquark violation one.}
\label{fig6}
\end{figure}
\begin{figure}[H]
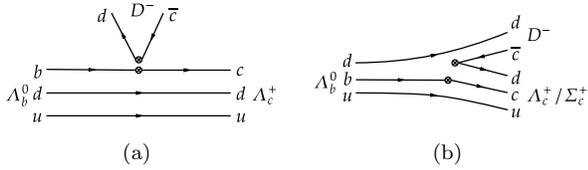

\centering
\subfigure[]{
\includegraphics[width=0.44\linewidth]{./image/fig2a.pdf}
}
\subfigure[]{
\includegraphics[width=0.44\linewidth]{./image/fig2b.pdf}
}
\caption{The Feynman diagrams of process $\Lambda_{b}^{0}\rightarrow\Lambda_{c}^{+}D^{-}$. The denotations of  subfigures (a) and (b) are the same as Fig.~\ref{fig6}.}
 \end{figure} 
 \begin{figure}[H]
\centering
\includegraphics[width=0.44\linewidth]{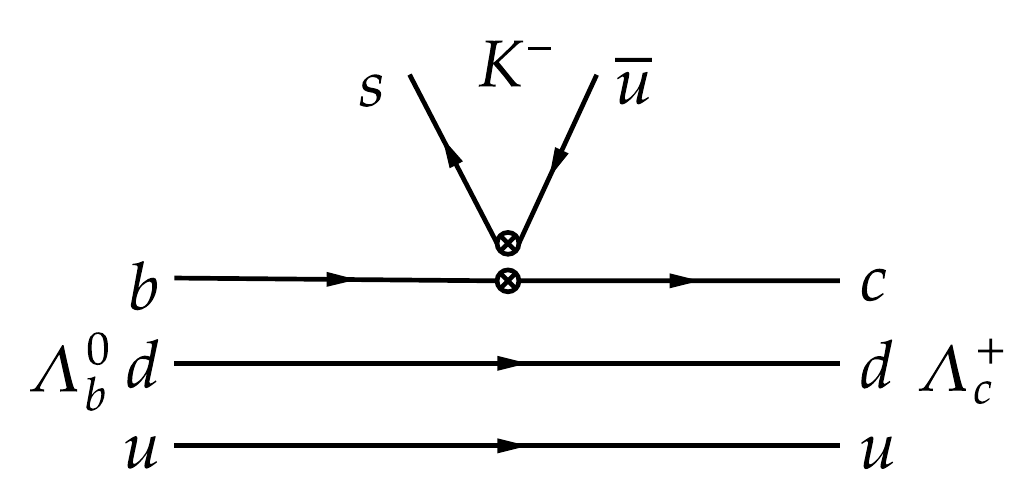}
\caption{The Feynman diagram of process $\Lambda_{b}^{0}\rightarrow \Lambda_{c}^{+}K^{-}$.}
 \end{figure}
\begin{figure}[H]
\centering
\includegraphics[width=0.44\linewidth]{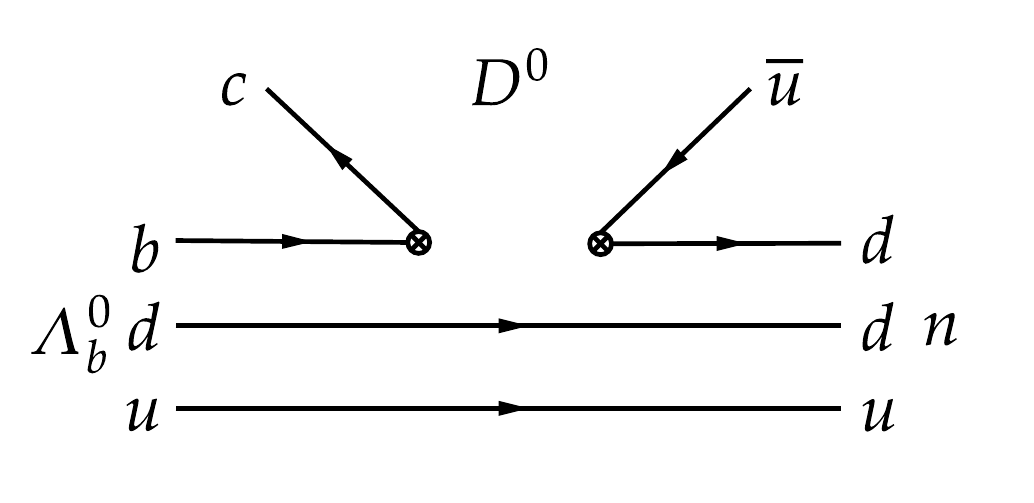}
\caption{The Feynman diagram of the color-suppressed process $\Lambda_{b}^{0}\rightarrow nD^{0}$.}
\label{fig8}
 \end{figure}
\begin{figure}[H]
\centering
\includegraphics[width=0.44\linewidth]{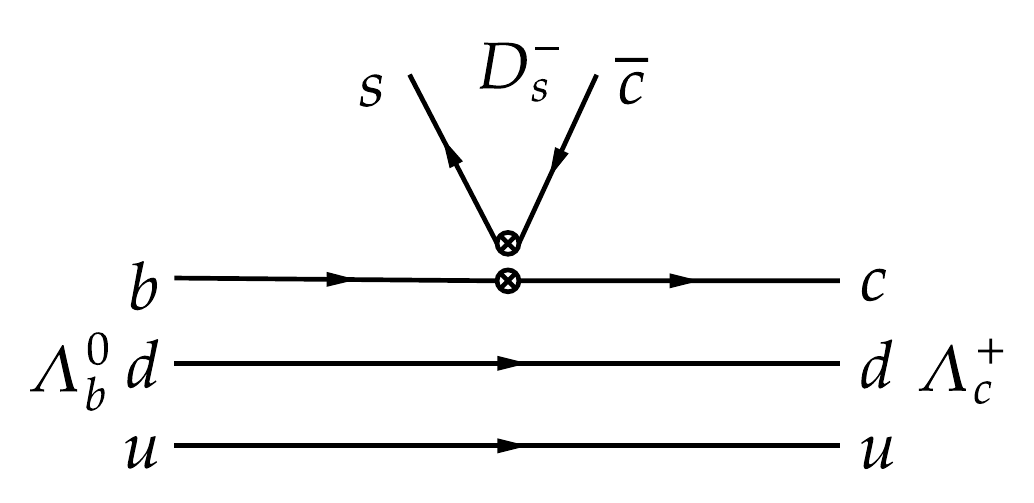}
\caption{The Feynman diagram of process $\Lambda_{b}^{0}\rightarrow \Lambda_{c}^{+}D_{s}^{-}$.}
 \end{figure} 
\begin{figure}[H]
\centering
\includegraphics[width=0.44\linewidth]{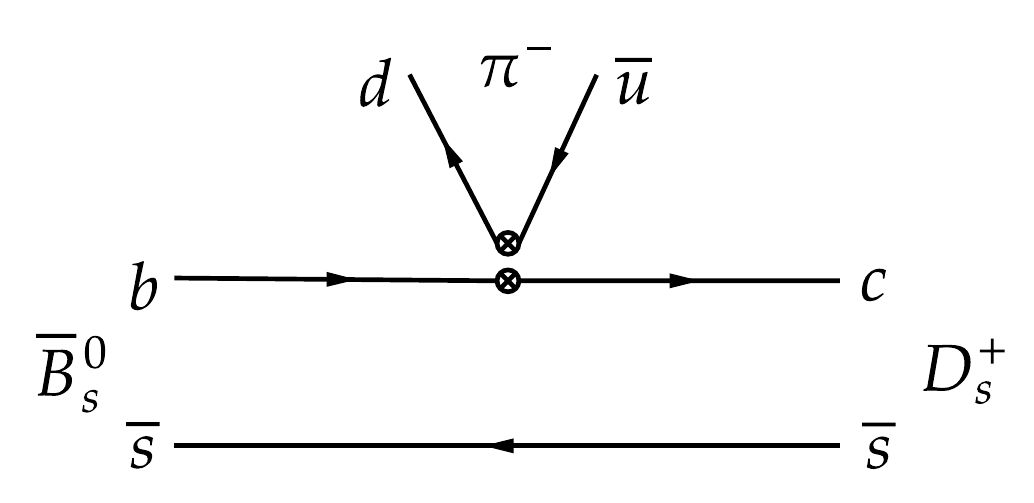}
\caption{The Feynman diagram of  process $\bar{B}_{s}^{0}\rightarrow D_{s}^{+}\pi^{-}$.}
 \end{figure}
\begin{figure}[H]
\centering
\includegraphics[width=0.44\linewidth]{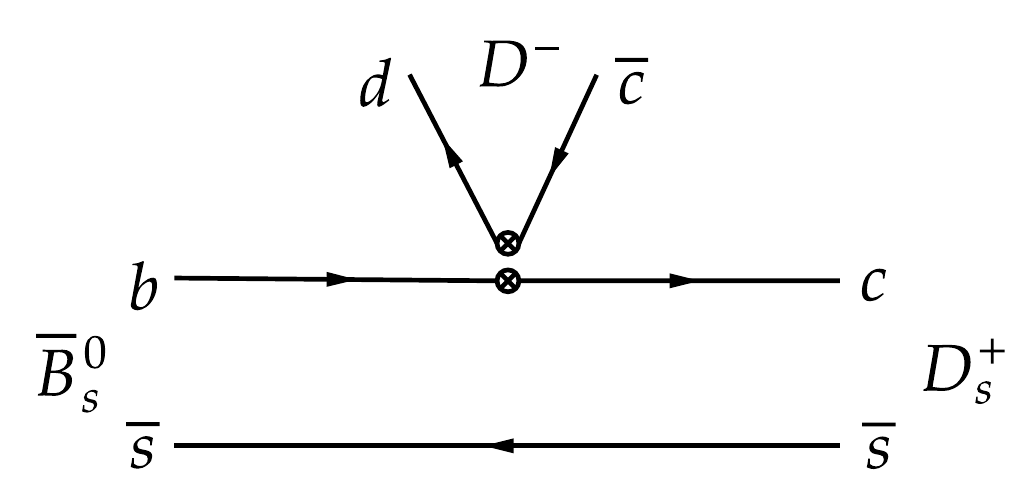}
\caption{The Feynman diagram of process $\bar{B}_{s}^{0}\rightarrow D_{s}^{+}D^{-}$.}
\end{figure}
\begin{figure}[H]
\centering
\includegraphics[width=0.44\linewidth]{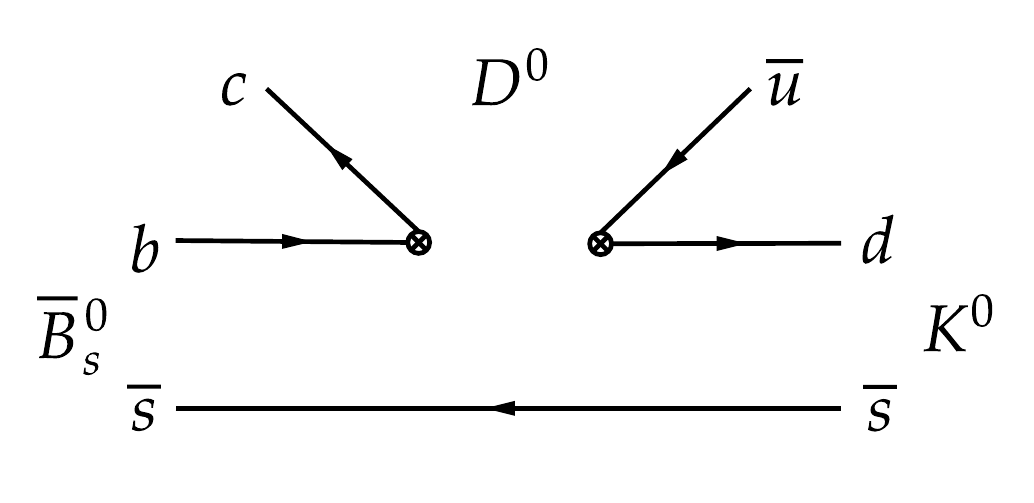}
\caption{The Feynman diagram of the color-suppressed process $\bar{B}_{s}^{0}\rightarrow K^{0}D^{0}$.}
 \end{figure}
\begin{figure}[H]
\centering
\subfigure[]{
\includegraphics[width=0.44\linewidth]{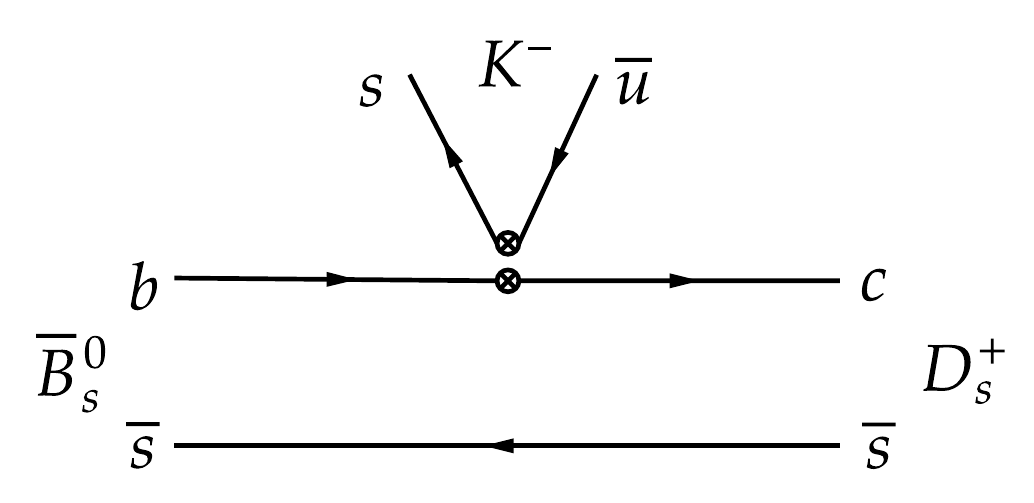}
}
\subfigure[]{
\includegraphics[width=0.44\linewidth]{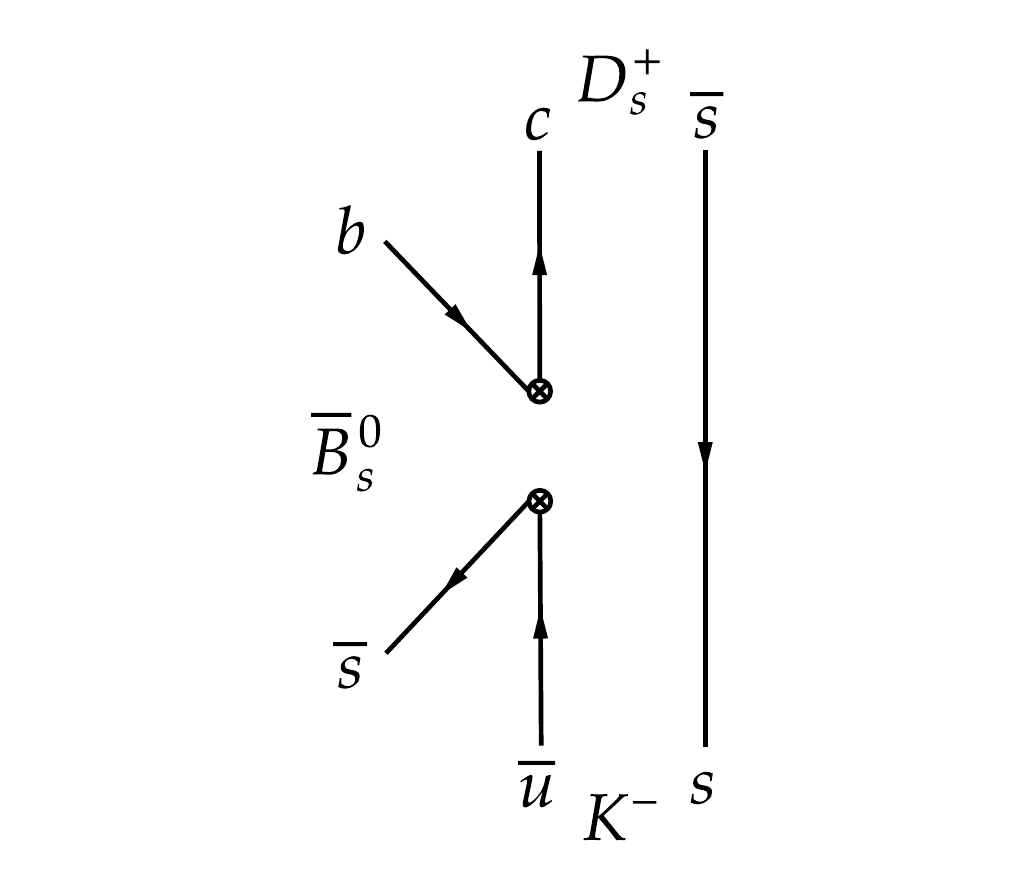}
}
\caption{The Feynman diagrams of process $\bar{B}_{s}^{0}\rightarrow D_{s}^{+}K^{-}$. The subfigure (b) shows the annihilation process.}
 \end{figure} 
\begin{figure}[H]
\centering
\subfigure[]{
\includegraphics[width=0.44\linewidth]{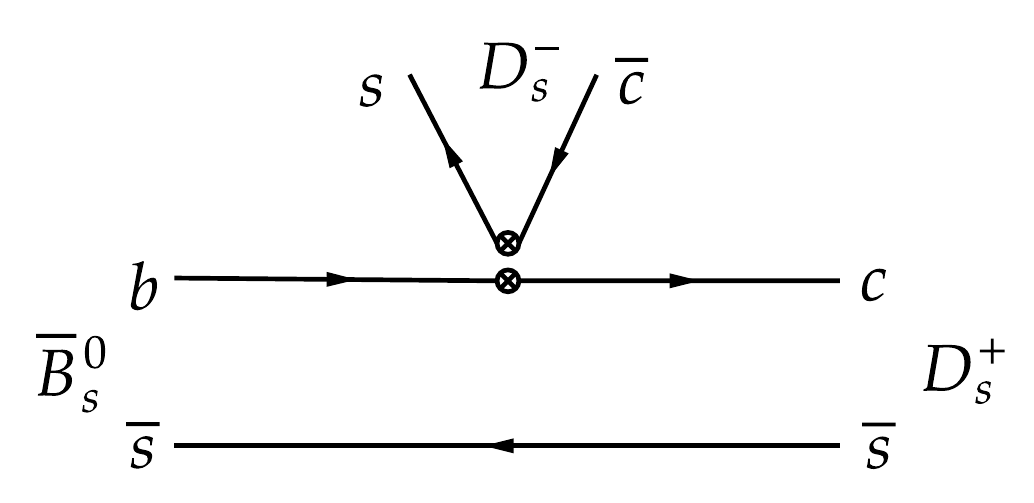}
}
\subfigure[]{
\includegraphics[width=0.44\linewidth]{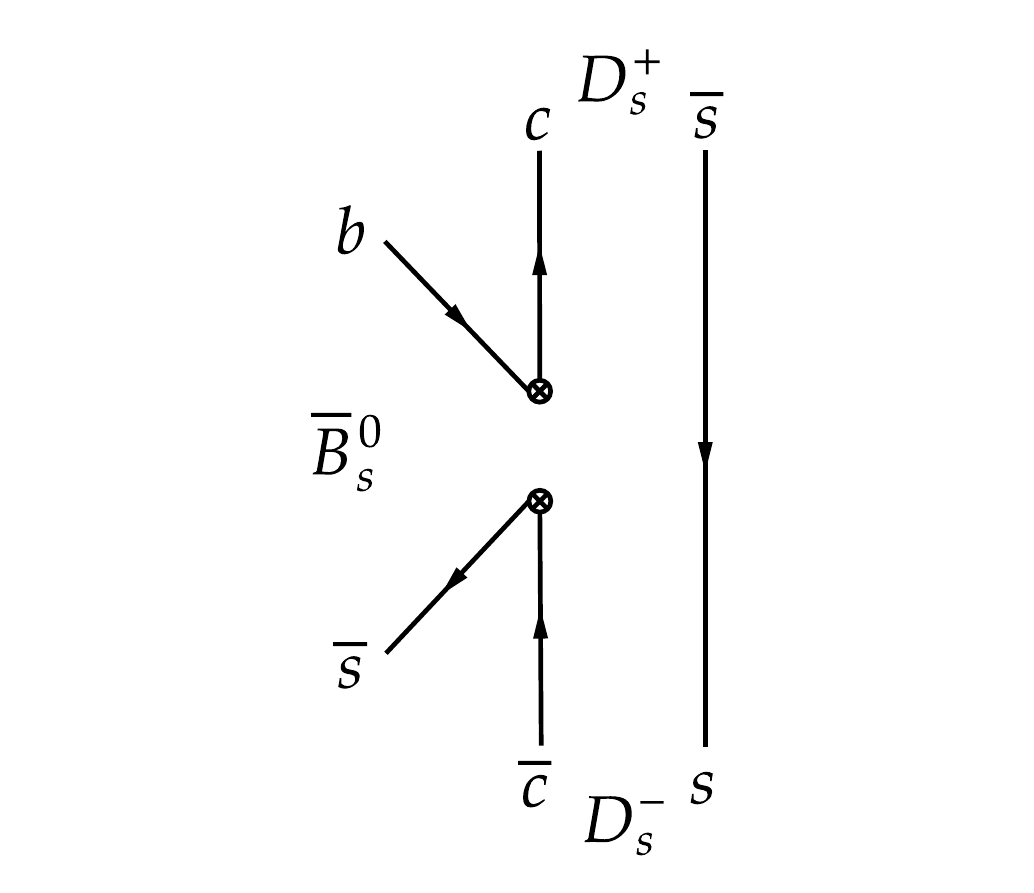}
}
\caption{The Feynman diagrams of  process $\bar{B}_{s}^{0}\rightarrow D_{s}^{+}D_{s}^{-}$. The subfigure (b) shows the annihilation process.}
\end{figure}
\begin{figure}[H]
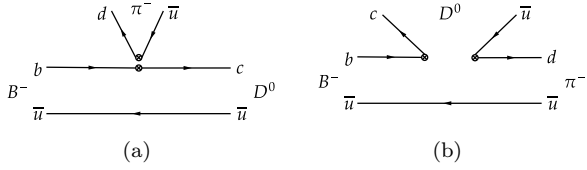

\centering
\subfigure[]{
\includegraphics[width=0.44\linewidth]{./image/fig5a.pdf}
}
\subfigure[]{
\includegraphics[width=0.44\linewidth]{./image/fig5b.pdf}
}
\caption{The Feynman diagrams of  process $B^{-}\rightarrow D^{0}\pi^{-}$. The subfigures (a) and (b) show the color-favored and color-suppressed processes, respectively.}
\label{fig15}
\end{figure}
\begin{figure}[H]
\centering
\subfigure[]{
\includegraphics[width=0.45\linewidth]{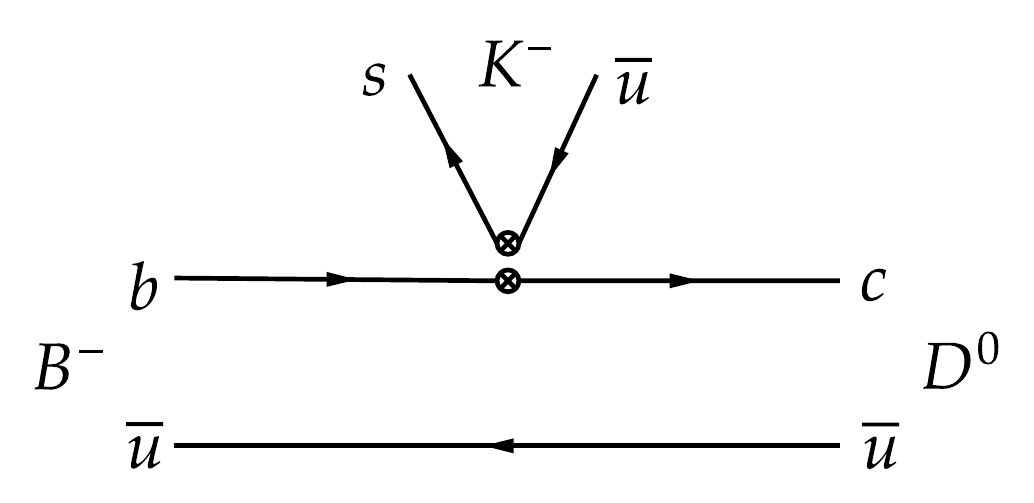}
}
\subfigure[]{
\includegraphics[width=0.45\linewidth]{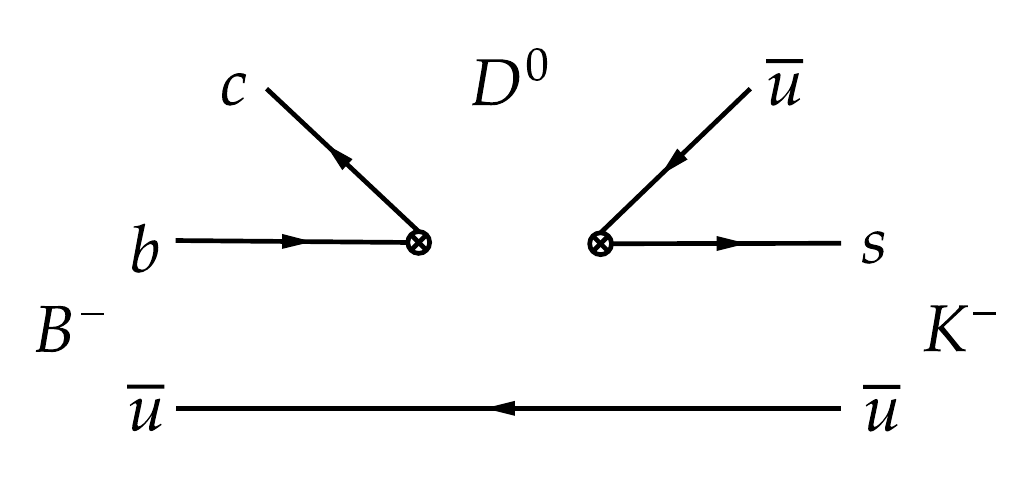}
}
\caption{The Feynman diagrams of  process $B^{-}\rightarrow D^{0}K^{-}$. The denotations of subfigures (a) and (b) are the same as Fig.~\ref{fig15}. }
\label{fig16}
\end{figure}
\begin{figure}[H]
\centering
\includegraphics[width=0.46\linewidth]{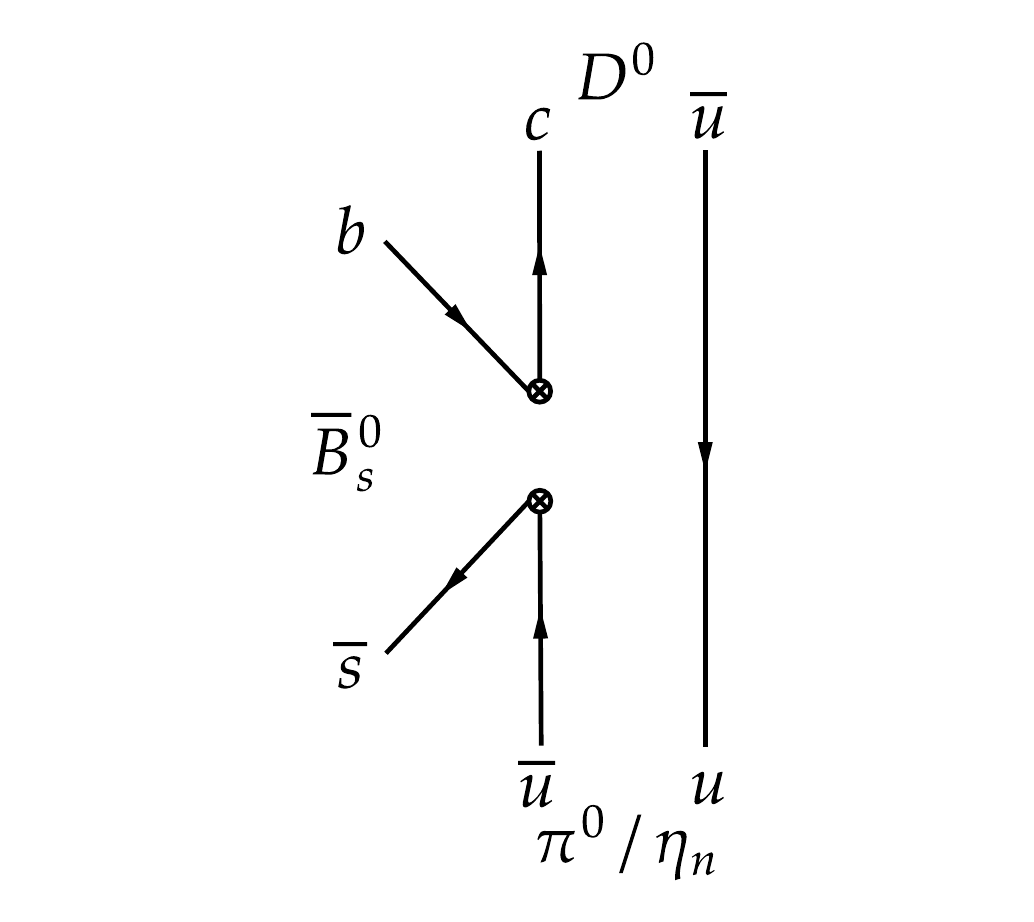}
\caption{The Feynman diagram of the annihilation  process $\bar{B}_{s}^{0}\rightarrow \pi^{0}D^{0}/\eta_{n}D^{0}$.}
\end{figure}
\begin{figure}[H]
\centering
\includegraphics[width=0.46\linewidth]{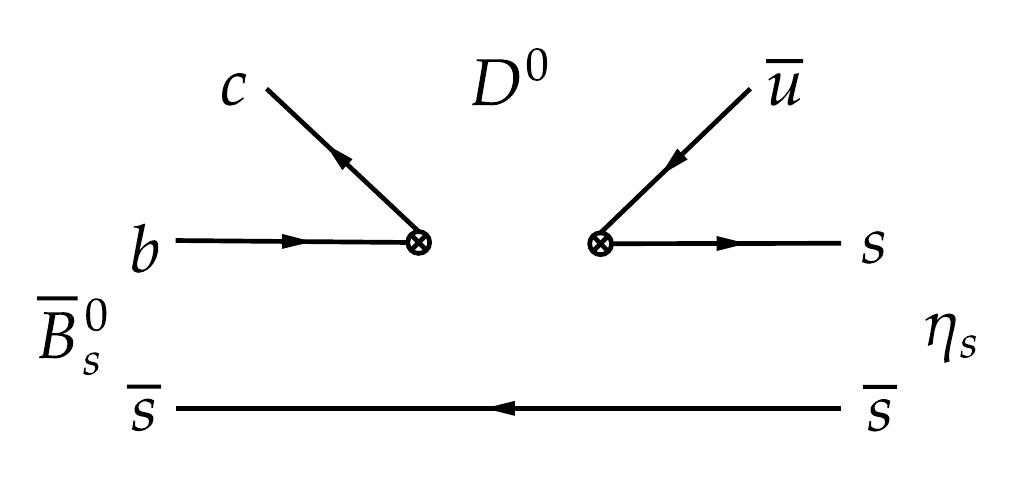}
\caption{The Feynman diagram of the color-suppressed process $\bar{B}_{s}^{0}\rightarrow \eta_{s}D^{0}$.}
\end{figure}
\vspace{1.5em}
\begin{figure}[H]
\centering
\includegraphics[width=0.46\linewidth]{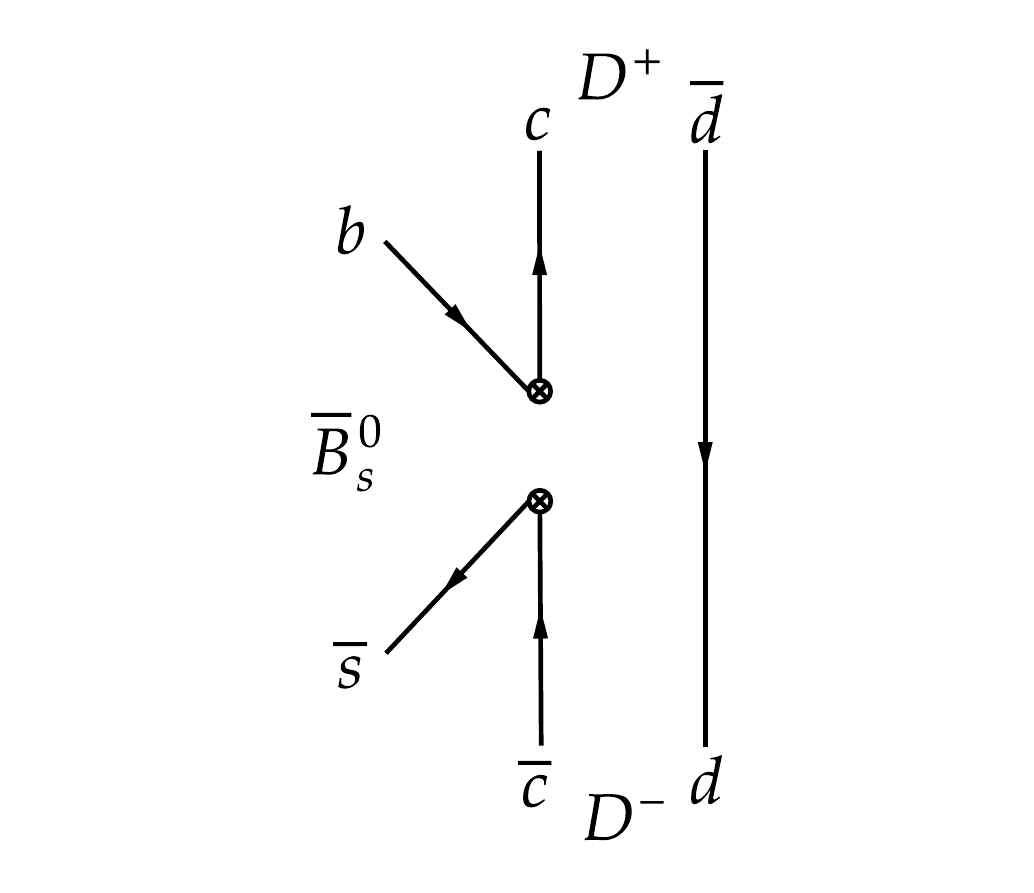}
\caption{The Feynman diagram of the annihilation  process $\bar{B}_{s}^{0}\rightarrow D^{+}D^{-}$.}
\end{figure}
\begin{figure}[H]
\centering
\includegraphics[width=0.46\linewidth]{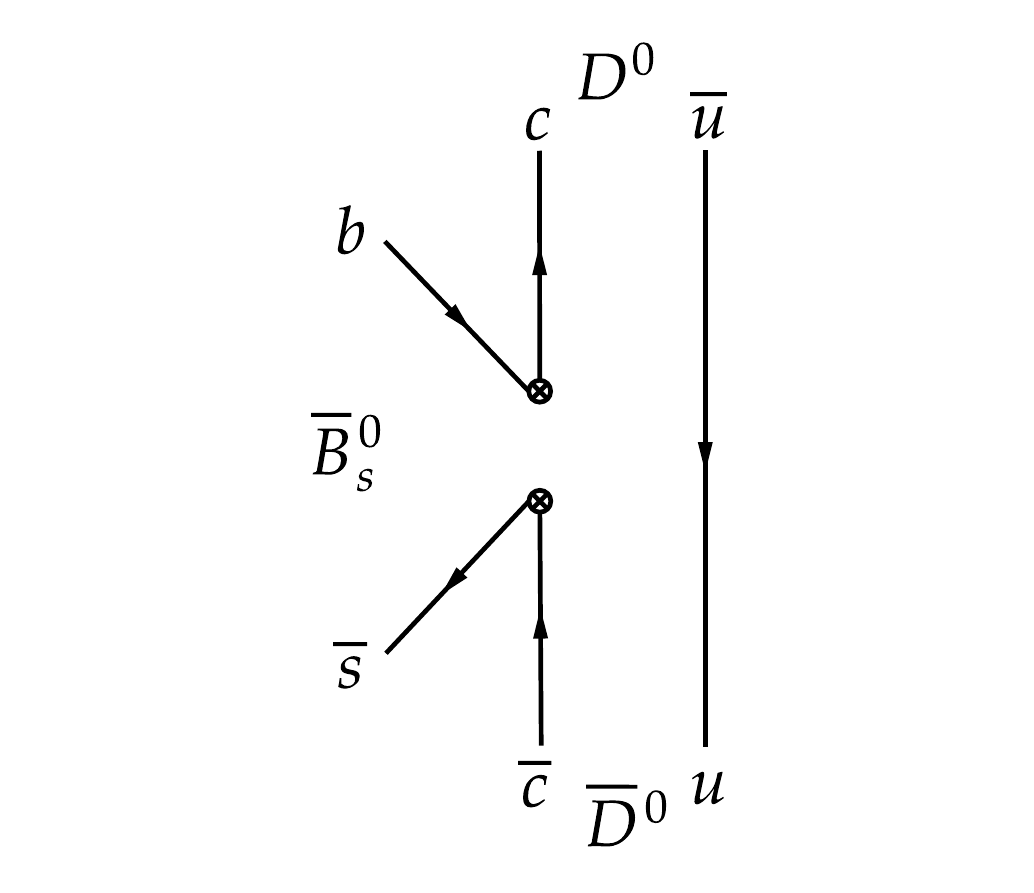}
\caption{The Feynman diagram of the annihilation  process $\bar{B}_{s}^{0}\rightarrow D^{0}\bar{D}^{0}$.}
\end{figure}
\vspace{2em}
 \section{Experiment Data}\label{AppendixA}
 Note that all the experiment data of decaying particles and branching ratio of each decay mode are taken from ~\cite{ParticleDataGroup:2022pth}.
 \begin{widetext}
 \begin{center}
\begin{table}[H]
    \centering
    \caption{The life time, decay widths and masses of the interested decaying particles.}
    \label{tab5:decaying_particle}
    \begin{tabular}{l|c|c|c}
    \hline\hline
  \textbf{Decaying particle} &\textbf{$\tau$[s]} &  \textbf{$\Gamma$($\frac{1}{\tau}$)[GeV]}   &   \textbf{$M$[GeV]}\\
    \hline
    \hline
   $\Lambda_b^0$  &$(1.471\pm0.009)\times10^{-12}$ &   $(4.481\pm0.027)\times10^{-13}$ &    $5.61960\pm0.00017$\\
   $\Xi_b^0$  &$(1.480\pm0.030)\times10^{-12}$ &   $(4.453\pm0.09)\times10^{-13}$ &    $5.79190\pm0.00050$\\
   $\Xi_b^-$  &$(1.572\pm0.040)\times10^{-12}$ &   $(4.193\pm0.107)\times10^{-13}$ &    $5.79700\pm0.00060$\\
   $\bar{B}_s^0$  &$(1.520\pm0.005)\times10^{-12}$ &   $(4.336\pm0.014)\times10^{-13}$ &    $5.36692\pm0.00010$\\
  $\bar{B}^0$  &$(1.519\pm0.004)\times10^{-12}$ &   $(4.339\pm0.011)\times10^{-13}$ &    $5.27966\pm0.00012$\\
  $B^-$  &$(1.638\pm0.004)\times10^{-12}$ &   $(4.024\pm0.010)\times10^{-13}$ &    $5.27934\pm0.00012$\\
   \hline
    \end{tabular}    
\end{table}
\begin{table}[H]
    \centering
    \caption{The branching fraction, partial decay width, momentum and phase space of each decay mode.}
    \label{tab6:phase_space}
    \begin{tabular}{l|c|c|c|c}
    \hline\hline
\textbf{Decay mode}  &  \textbf{Fraction($\frac{\Gamma_i}{\Gamma}$)}  &  \textbf{$\Gamma_{i}$[GeV]} & \textbf{$p$[GeV]} & \textbf{$8\pi^{4}\Phi_2[1]$}\\
    \hline
    \hline
     $\Lambda_{b}^{0}\rightarrow \Lambda_c^{+}\pi^{-}$\textbf{(Input)}  &   $(4.9\pm0.4)\times10^{-3}$ &   $(2.20 \pm0.18)\times10^{-15}$ & $2.342$ & $0.017$\\
  $\Lambda_{b}^{0}\rightarrow \Lambda_c^{+}D^{-}$\textbf{(Input)}  &   $(4.6\pm0.6)\times10^{-4}$ &    $(2.06\pm0.27)\times10^{-16}$ & $1.886$ & $0.013$\\
   $\bar{B}_{s}^{0}\rightarrow D^{+}_{s}\pi^{-}$\textbf{(Input)}  &   $(2.98\pm0.14)\times10^{-3}$ & $(1.30\pm0.06)\times10^{-15}$ &$2.320$ &  $0.017$\\
   $\bar{B}_{s}^{0}\rightarrow D^{0}K^{0}$\textbf{(Input)} &$(4.3\pm0.9)\times10^{-4}$ & $(1.86\pm0.39)\times10^{-16}$ & $2.330$ &$0.017$\\
   $\bar{B}_{s}^{0}\rightarrow D^{+}_{s}D^{-}$\textbf{(Input)}   &   $(2.8\pm0.5)\times10^{-4}$ & $(1.21\pm0.22)\times10^{-16}$ &$1.875$ &  $0.014$\\
   $\Lambda_{b}^{0}\rightarrow \Lambda_{c}^{+}K^{-}$   &   $(3.56\pm0.28)\times10^{-4}$ &   $(1.60\pm0.13)\times10^{-16}$ & $2.314$ & $0.016$\\
   $\Lambda_{b}^{0}\rightarrow \Lambda_{c}^{+}D_{s}^{-}$\textbf{(Input)}   &   $(1.10\pm0.10)\%$ &    $(4.93\pm0.45)\times10^{-15}$ & $1.833$ & $0.013$\\
    $\bar{B}_{s}^{0}\rightarrow D_{s}^{+}K^{-}$   &   $(2.25\pm0.12)\times10^{-4}$ &    $(9.76\pm0.52)\times10^{-17} $& $2.293$ & $0.017$\\
   $\bar{B}_{s}^{0}\rightarrow D_{s}^{+}D_{s}^{-}$   &   $(4.4\pm0.5)\times10^{-3}$ &    $(1.91\pm0.22)\times10^{-15}$ & $1.824$ & $0.014$\\
   \hline
    \end{tabular}     
\end{table}
\end{center} 
\begin{center}
\begin{table}[H]
    \centering
    \caption {The errors brought by $V_{bc}$, $V_{qq'}$, $a$, $c$, $d$ and $F$ corresponding to each decay mode.(Units:GeV)}
    \label{tab7}
    \begin{tabular}{l|c|c|c|c|c|c|c}
    \hline
    \hline
\textbf{Decay mode} & \textbf{$V_{bc}$ error} &\textbf{$V_{qq'}$ error}&\textbf{$a$ error}&\textbf{$c$ error}&\textbf{$d$ error}&\textbf{$F$ error}&\textbf{Total error} \\
    \hline
    \hline   $\Lambda_{b}^{0}\rightarrow\Sigma_{c}^{+}\pi^{-}$ &$1.81\times10^{-16}$ &$1.68\times10^{-18}$ &$-$ &$8.13\times10^{-17}$ &$4.90\times10^{-16}$ &$-$ &$5.29\times10^{-16}$\\
    $\Lambda_{b}^{0}\rightarrow nD^{0}$ &$1.20\times10^{-17}$ &$1.11\times10^{-19}$ &$7.10\times10^{-18}$ &$-$ &$-$ &$3.70\times10^{-17}$  &$3.93\times10^{-17}$\\ $\Lambda_{b}^{0}\rightarrow\Sigma_{c}^{+}D^{-}$  &$1.10\times10^{-17}$ &$5.79\times10^{-18}$ &$-$ &$4.94\times10^{-18}$ &$2.98\times10^{-17}$ &$-$  &$3.26\times10^{-17}$\\
    $\Lambda_b^0\rightarrow \Lambda_c^+K^-$ &$6.39\times10^{-18}$ &$6.64\times10^{-19}$ &$3.78\times10^{-18}$ &$-$ &$-$ &$-$ &$7.45\times10^{-18}$\\
    $\bar{B}_{s}^{0}\rightarrow D_{s}^{+}K^{-}$  &$6.94\times10^{-18}$ &$7.22\times10^{-19}$ &$4.10\times10^{-18}$ &$-$ &$-$ &$-$
    &$8.09\times10^{-18}$\\    
    $\bar{B}_{s}^{0}\rightarrow D_{s}^{+}D_{s}^{-}$  &$2.84\times10^{-16}$ &$5.09\times10^{-17}$ &$1.68\times10^{-16}$ &$-$ &$-$ &$-$ 
    &$3.34\times10^{-16}$\\ 
    \hline
    $\Xi_{b}^{0}\rightarrow\Xi_{c}^{+}\pi^{-}$   &$8.06\times10^{-17}$ &$7.47\times10^{-19}$  &$4.76\times10^{-17}$
    &$-$  &$-$ &$-$ 
    &$9.36\times10^{-17}$\\
    $\Xi_{b}^{0}\rightarrow\Xi_{c}^{+}K^{-}$    &$6.08\times10^{-18}$ &$6.32\times10^{-19}$  &$3.59\times10^{-18}$
    &$-$  &$-$ &$-$ 
    &$7.09\times10^{-18}$\\
    $\Xi_{b}^{0}\rightarrow\Xi_{c}^{+}D^{-}$     &$8.14\times10^{-18}$ &$4.29\times10^{-18}$ &$4.81\times10^{-18}$
    &$-$   &$-$ &$-$ 
    &$1.04\times10^{-17}$\\
    $\Xi_{b}^{0}\rightarrow\Xi_{c}^{+}D_{s}^{-}$  &$2.48\times10^{-16}$
    &$4.45\times10^{-17}$ &$1.47\times10^{-16}$
    &$-$   &$-$ &$-$ 
    &$2.92\times10^{-16}$\\
    $\bar{B}^{0}\rightarrow D^{+}\pi^{-}$  &$9.44\times10^{-17}$ &$8.76\times10^{-19}$ &$5.58\times10^{-17}$
    &$-$   &$-$ &$-$ 
    &$1.10\times10^{-16}$\\
   $\bar{B}^{0}\rightarrow D^{+}K^{-}$  &$7.13\times10^{-18}$ &$7.41\times10^{-19}$ &$4.21\times10^{-18}$
   &$-$   &$-$  &$-$ 
   &$8.31\times10^{-18}$\\
   $\bar{B}^{0}\rightarrow D^{+}D^{-}$  &$9.56\times10^{-18}$ &$5.04\times10^{-18}$ &$5.65\times10^{-18}$
   &$-$   &$-$  &$-$ 
   &$1.22\times10^{-17}$\\
   $\bar{B}^{0}\rightarrow D^{+}D_{s}^{-}$  &$2.92\times10^{-16}$ &$5.23\times10^{-17}$ &$1.72\times10^{-16}$
   &$-$   &$-$  &$-$ 
   &$3.43\times10^{-16}$\\
   \hline
    $\Xi_{b}^{-}\rightarrow\Xi_{c}^{0}\pi^{-}$   &$4.86\times10^{-16}$ &$4.51\times10^{-18}$  &$6.27\times10^{-17}$
    &$6.59\times10^{-16}$   
    &$9.61\times10^{-16}$ &$-$ 
    &$1.26\times10^{-15}$\\
    $\Xi_{b}^{-}\rightarrow\Xi_{c}^{0}K^{-}$  
    &$6.08\times10^{-18}$ &$6.32\times10^{-19}$  &$3.59\times10^{-18}$
    &$-$   &$-$ &$-$ 
    &$7.09\times10^{-18}$\\
    $\Xi_{b}^{-}\rightarrow\Xi_{c}^{0}D^{-}$  
    &$2.62\times10^{-17}$ &$1.38\times10^{-17}$ &$5.79\times10^{-18}$
    &$3.90\times10^{-17}$   
    &$3.99\times10^{-17}$
    &$-$ 
    &$6.34\times10^{-17}$\\
    $\Xi_{b}^{-}\rightarrow\Xi_{c}^{0}D_{s}^{-}$  &$2.48\times10^{-16}$ &$4.45\times10^{-17}$ &$1.46\times10^{-16}$
    &$-$   &$-$ &$-$ 
    &$2.91\times10^{-16}$\\   
   $B^{-}\rightarrow D^{0}\pi^{-}$  &$1.78\times10^{-16}$ &$1.65\times10^{-18}$ &$1.05\times10^{-16}$
   &$-$   
   &$-$
   &$1.47\times10^{-16}$
   &$2.54\times10^{-16}$\\
   $B^{-}\rightarrow D^{0}K^{-}$ &$1.22\times10^{-17}$ &$1.27\times10^{-18}$ &$7.23\times10^{-18}$
   &$-$
   &$-$
   &$8.85\times10^{-18}$
   &$1.68\times10^{-17}$\\
   $B^{-}\rightarrow D^{0}D^{-}$  &$9.58\times10^{-18}$ &$5.05\times10^{-18}$ &$5.66\times10^{-18}$
   &$-$   &$-$ &$-$
   &$1.22\times10^{-17}$\\
   $B^{-}\rightarrow D^{0}D_{s}^{-}$  &$2.92\times10^{-16}$ &$5.24\times10^{-17}$ &$1.73\times10^{-16}$
   &$-$   &$-$ &$-$
   &$3.43\times10^{-16}$\\
   \hline
   \end{tabular}
   \label{tab:amplitude}    
\end{table}
\end{center} 
\end{widetext}
\section{The Treatment of Error} \label{AppendixC}
In this work, the errors are brought by the corresponding CKM matrix elements, fitted parameters $a$, $c$, $d$, and $F$, while the errors brought by the masses of particles, momentum $p$ and Fermi coupling constant $G_{F}$ are neglected due to their small contributions. The details are shown in Tab.~\ref{tab:amplitude}.

\end{document}